\documentclass[aps,prc,twocolumn,showpacs,floatfix,nofootinbib,
preprintnumbers,superscriptaddress,amsmath,amssymb]{revtex4-1}

\usepackage{CJKutf8}
\usepackage{epsfig}
\usepackage{color}
\usepackage{rotating}
\usepackage{natbib}
\usepackage{verbatim}
\usepackage{epstopdf}
\usepackage{graphicx}
\usepackage{changes}

\renewcommand{\vec}[1]{\mbox{\boldmath $#1$}}

\begin{document}

\begin{CJK*}{UTF8}{gbsn}

\title{Nucleon localization and fragment formation in nuclear fission}

\author{C. L. Zhang (张春莉)}
\affiliation
  {
     NSCL/FRIB Laboratory, Michigan State University, East Lansing, Michigan 48824, USA
  }

\author{B. Schuetrumpf}
\affiliation
  {
     NSCL/FRIB Laboratory, Michigan State University, East Lansing, Michigan 48824, USA
  }

\author{W. Nazarewicz}
\affiliation
  {Department of Physics and Astronomy and FRIB Laboratory, Michigan State University, East Lansing, Michigan 48824, USA}
\affiliation{Institute of Theoretical Physics, Faculty of Physics, 
University of Warsaw, 02-093 Warsaw, Poland}

\date{\today}

\begin{abstract}
\begin{description}

\item[Background]  An electron localization measure was originally introduced to characterize chemical bond structures in molecules. Recently, a nucleon localization based on Hartree-Fock densities has been introduced to investigate $\alpha$-cluster structures in light nuclei. Compared to the local nucleonic densities, the nucleon localization function has been shown to be an excellent indicator of shell effects and cluster correlations.

\item[Purpose] Using the spatial nucleon localization measure, we investigate the emergence of fragments in fissioning heavy nuclei.

\item[Methods] To illustrate basic concepts of nucleon localization, we employ the self-consistent energy density functional method with a quantified energy density functional optimized for fission  studies. 

\item[Results] We study the particle densities and spatial nucleon localization distributions along the fission pathways of $^{264}$Fm, $^{232}$Th and $^{240}$Pu.  We  demonstrate  that the fission fragments  are formed fairly early in the evolution, well before scission. We illustrate the usefulness of the localization measure by showing how the hyperdeformed state of $^{232}$Th can be understood in terms of a quasimolecular state made of $^{132}$Sn and $^{100}$Zr fragments.

\item[Conclusions] 
Compared to nucleonic distributions, the nucleon localization function more effectively quantifies nucleonic clustering: its characteristic oscillating pattern, traced back to shell effects, is a clear fingerprint of cluster/fragment configurations. This is of particular interest for studies of fragment formation and fragment identification in fissioning nuclei.
\end{description}
\end{abstract}


\maketitle

\end{CJK*}
\section{Introduction}

The appearance of cluster states in atomic nuclei  is a ubiquitous phenomenon with many implications for both  nuclear physics and astrophysics \cite{beck2010clusters, beck2012clusters, beck2014clusters,vonOertzen2006,DelionCluster}. While several factors are known to contribute to clustering, a comprehensive microscopic  understanding of this phenomenon still remains elusive. Cluster configurations can be energetically favorable due to the large binding energy per nucleon in constituent clusters, such as $\alpha$ particles.  The binding-energy argument has often been used to explain properties of $\alpha$-conjugate nuclei \cite{Hafstad}, cluster emission \cite{Rose1984,Aleksandrov1984} and fission \cite{BLynn}, and  the appearance of a gas of light clusters in low-density nuclear matter \cite{Ropke2006,Girod13,Ebran14a} and in  the interior region of heavy nuclei \cite{BRINK1973109}. Another important factor is the coupling to decay channels; this explains \cite{Okolowicz12,Okolowicz13} the very occurrence of cluster states  at low excitation energies around  cluster-decay thresholds \cite{Ikeda68}.

The microscopic description of cluster states requires the use of an advanced many-body, open-system framework \cite{Okolowicz12,Okolowicz13,Cluster12} employing realistic interactions, and there has been significant progress in this area \cite{Yoshida13,Epelbaum12,Epelbaum14,Elh15,QMC}. For  a global characterization of cluster states in nuclei, a good starting point is density functional theory \cite{Jones15} based on a realistic nuclear energy density functional, or its self-consistent mean-field variant with density-dependent effective interactions \cite{bender2003self}, to which we shall refer as the energy density functional method (EDFM) in the following. Within EDFM, cluster states have a simple interpretation in terms of quasimolecular states. Since the mean-field approach is rooted in the variational principle, the binding-energy argument favors clustering in certain configurations characterized by large shell effects of constituent fragments \cite{Leander,Flocard84,Marsh86,Freer95,maruhn2010linear,Ichikawa11,ebran2012atomic,ebran2014density}; the characteristics of cluster states can be indeed traced back to both the symmetries and geometry of the nuclear mean-field \cite{Hecht77,Nazarewicz1992}. 

The degree of clustering in nuclei is difficult to assess quantitatively in EDFM as the single particle (s.p.) wave functions are spread throughout the nuclear volume; hence, the resulting nucleonic distributions are rather crude indicators of cluster structures as their behavior in the nuclear interior is fairly uniform. Therefore, in this study, we utilize a different measure called spatial localization, which is a more selective signature of clustering and cluster shell structure. The localization, originally introduced for the identification of localized electronic groups in atomic and molecular systems \cite{Becke1990,savin1997elf,scemama2004electron,Kohout04,burnus2005time,Poater}, has recently been applied to characterize clusters in light nuclei \cite{Reinhard2011}. In this work, we investigate the usefulness of the spatial localization as a tool to identify fission fragments in heavy fissioning nuclei.

This article is organized as follows: Section~\ref{model} gives a brief introduction
to the EDFM and the localization measure employed in this work. The results for fissioning  nuclei $^{264}$Fm, $^{232}$Th,  and $^{240}$Pu are presented in Sec.~\ref{sec:heavynuclei-HFB}. Finally, the summary and outlook are given in Sec.~\ref{summary}.

\section{Model}
\label{model}

\subsection{EDFM Implementation}

In superfluid nuclear EDFM, the binding energy is expressed through the general density matrix \cite{ring2004nuclear,bender2003self}. By applying the variational principle to s.p. wave functions (Kohn-Sham orbitals), the self-consistent  Hartree-Fock-Bogoliubov (HFB) equations are derived. Nuclear EDFM has been successfully used to describe properties of ground states and selected collective  states across the nuclear landscape \cite{bender2003self,bogner2013,erler12,Agbemava14}.

In this work, we use Skyrme energy density functionals which are expressed in terms of local nucleonic densities and currents. We employ the UNEDF1 functional optimized for fission \cite{kortelainen2012nuclear} in the presence of pairing treated by means of the Lipkin-Nogami approximation as in Ref.~\cite{Sto03}. 
The constrained  HFB equations are solved 
by using the symmetry-unconstrained code  HFODD \cite{schunck2012solution}.

\subsection{Spatial Localization}
\label{model:localization}
The spatial localization measure was originally introduced in atomic and molecular physics to characterize chemical bonds in electronic systems. It also turned out to be useful for visualizing  cluster structures in light nuclei \cite{Reinhard2011}. It can be derived by considering the conditional probability of finding a nucleon within a distance $\delta$ from a given nucleon
at $\vec{r}$ with the same spin $\sigma$ ($=\uparrow$ or $\downarrow$)  and isospin $q$ ($=n$ or $p$). As discussed in \cite{Becke1990,Reinhard2011}, the expansion of this probability with respect to $\delta$ can be written as
\begin{equation}\label{eqn:probability}
R_{q\sigma}(\vec{r},\delta)\approx{1\over 3}\left(\tau_{q\sigma}-{1\over 4}\frac{|\vec{\nabla}\rho_{q\sigma}|^2}{\rho_{q\sigma}}-\frac{\vec{j}^2_{q\sigma}}{\rho_{q\sigma}}\right)\delta^2+\mathcal{O}(\delta^3),
\end{equation} 
where $\rho_{q\sigma}$, $\tau_{q\sigma}$, $\vec{j}_{q\sigma}$, and $\vec{\nabla}\rho_{q\sigma}$ are the particle density, kinetic energy density, current density, and density gradient, respectively. They can be expressed through the canonical HFB orbits $\psi_\alpha(\vec{r}\sigma)$:
\begin{subequations}
\begin{eqnarray}
\rho_{q\sigma}(\vec{r})&=&\sum_{\alpha\in q}v^2_{\alpha}|\psi_\alpha(\vec{r}\sigma)|^2,\\
\tau_{q\sigma}(\vec{r})&=&\sum_{\alpha\in q}v^2_{\alpha}|\vec{\nabla}\psi_\alpha(\vec{r}\sigma)|^2,\\
\vec{j}_{q\sigma}(\vec{r})&=&\sum_{\alpha\in q}v^2_{\alpha}\mathrm{Im}[\psi^*_\alpha(\vec{r}\sigma)\vec{\nabla}\psi_\alpha(\vec{r}\sigma)],
\\
\vec{\nabla}\rho_{q\sigma}(\vec{r})&=&2\sum_{\alpha\in q}v^2_{\alpha}\mathrm{Re}[\psi^*_\alpha(\vec{r}\sigma)\vec{\nabla}\psi_\alpha(\vec{r}\sigma)],
\end{eqnarray}
\end{subequations}
with $v^2_{\alpha}$ being the canonical occupation probability. Thus, the expression in the parentheses of Eq.~(\ref{eqn:probability}) can serve as a localization measure.
Unfortunately, this expression is neither dimensionless nor normalized. A natural choice for normalization is the Thomas-Fermi kinetic energy density $\tau^\mathrm{TF}_{q\sigma}={3\over 5}(6\pi^2)^{2/3}\rho_{q\sigma}^{5/3}$.
Considering that the spatial localization and  $R_{q\sigma}$ are in an inverse relationship, a dimensionless and normalized expression for the localization measure can be written as
\begin{equation} \label{eqn:localization}
\mathcal{C}_{q\sigma}(\vec{r})=\left[1+\left(\frac{\tau_{q\sigma}\rho_{q\sigma}-{1\over 4}|\vec{\nabla}\rho_{q\sigma}|^2-\vec{j}^2_{q\sigma}}{\rho_{q\sigma}\tau^\mathrm{TF}_{q\sigma}}\right)^2\right]^{-1}.
\end{equation}
We note that the combination $\tau_{q\sigma}\rho_{q\sigma}-\vec{j}^2_{q\sigma}$
guarantees the Galilean invariance of the localization measure \cite{Engel75}. In this work, time reversal symmetry is conserved; hence, $\vec{j}_{q\sigma}$ vanishes.

A  value of $\mathcal{C}$ close to one  indicates that the probability of finding two nucleons with the same spin and isospin at the same spatial location is very low. Thus the nucleon's localization is large at that point. In particular, nucleons making up the alpha particle are perfectly localized \cite{Reinhard2011}. Another interesting case is $\mathcal{C}=1/2$, which corresponds to a  homogeneous Fermi gas as found in nuclear matter. When applied to many-electron systems, the quantity  $\mathcal{C}$ is referred to as the electron localization function, or ELF. In nuclear applications, the measure of localization (\ref{eqn:localization}) shall thus be called the nucleon localization function (NLF).

The above definition of the NLF works well in regions with non-zero nucleonic density. When the local densities become very small in the regions outside the range of the nuclear mean field, numerical instabilities can appear. On the other hand, when the particle density is close to zero, localization is no longer a meaningful quantity. Consequently, when visualizing localizations for finite nuclei in the 2D plots shown in this paper, we multiply the NLF by a normalized particle density $\mathcal{C}(\vec{r})\rightarrow\mathcal{C}(\vec{r})\rho_{q\sigma}(\vec{r})/[\mathrm{max}(\rho_{q\sigma}(\vec{r})]$. 

\section{Localization in fissioning heavy nuclei}
\label{sec:heavynuclei-HFB}

Based on the examples shown in Ref.~\cite{Reinhard2011}, we know that the oscillating pattern of  NLFs is an excellent tool for visualizing cluster structures in light nuclei. In this work, we apply this tool to monitor the development and evolution of fission fragments in $^{264}$Fm, $^{232}$Th, and $^{240}$Pu. 

We begin from  the discussion of the symmetric fission of $^{264}$Fm, a subject of several recent DFT studies 
\cite{Staszczak09,Sadhukhan14,Simenel14,Zhao15}. As shown in Ref.~\cite{Staszczak09}, at large values of the mass quadrupole moment $Q_{20}$, the static fission pathway 
of $^{264}$Fm is symmetric, with a neck emerging at  $Q_{20}\approx 145$\,b, and the scission point  reached at $Q_{20} \approx 265$\,b, above which $^{264}$Fm splits into two fragments. 
The appearance of the static symmetric fission  pathway in $^{264}$Fm is due to shell 
effects in the emerging fission fragments associated with  the doubly magic nucleus $^{132}$Sn \cite{Hulet89}.

\begin{figure}[htb]
   \includegraphics[width=\linewidth]{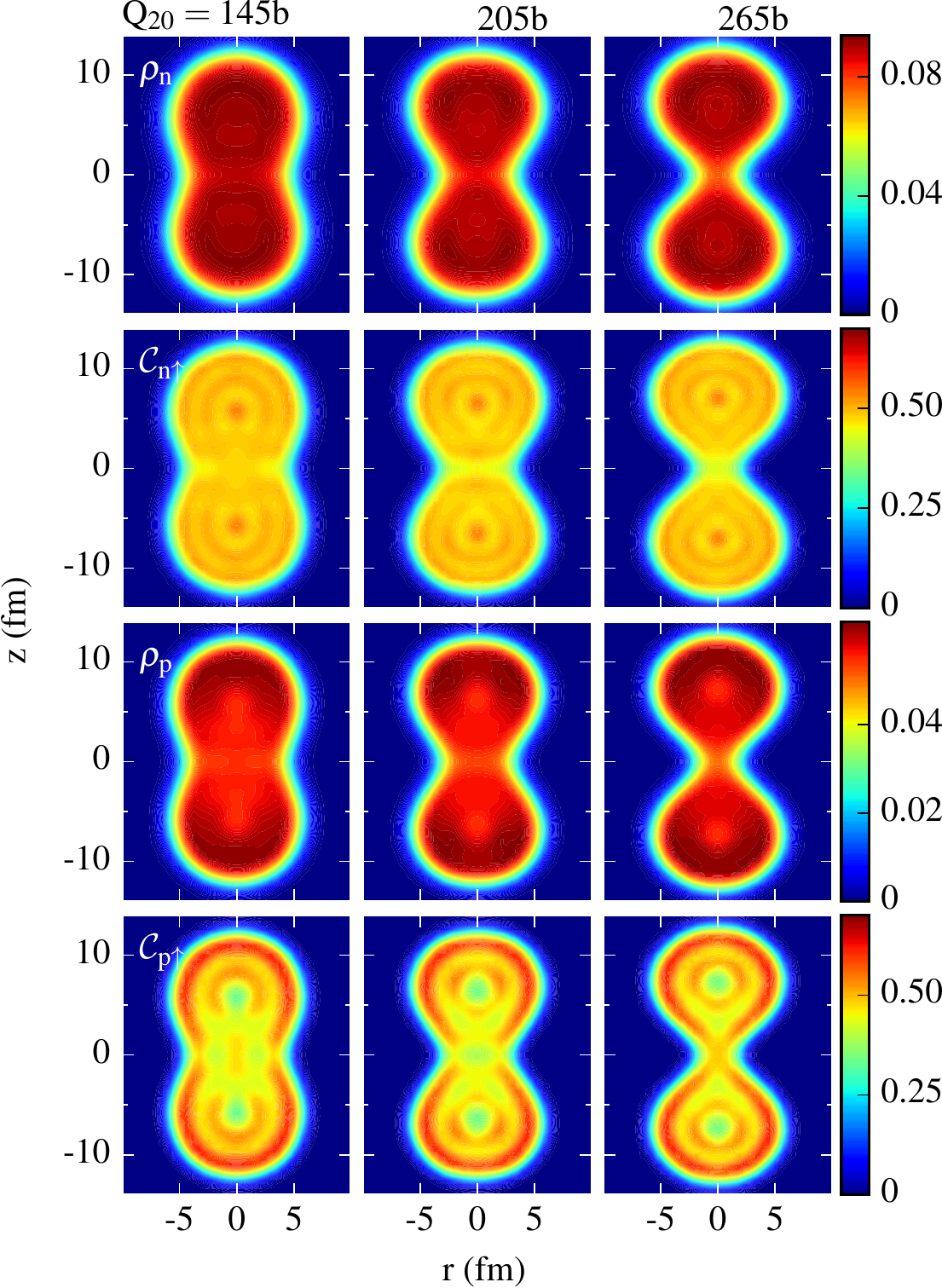} 
   \caption{Nucleonic densities (in nucleons/fm$^3$) and spatial localizations for $^{264}$Fm obtained from HFB calculations with UNEDF1 for three configurations along the symmetric fission pathway corresponding to different values of the mass quadrupole moment $Q_{20}$ and decreasing neck size.}
   \label{fig:fm264}
\end{figure}

\begin{figure}[htb]
   \includegraphics[width=\linewidth]{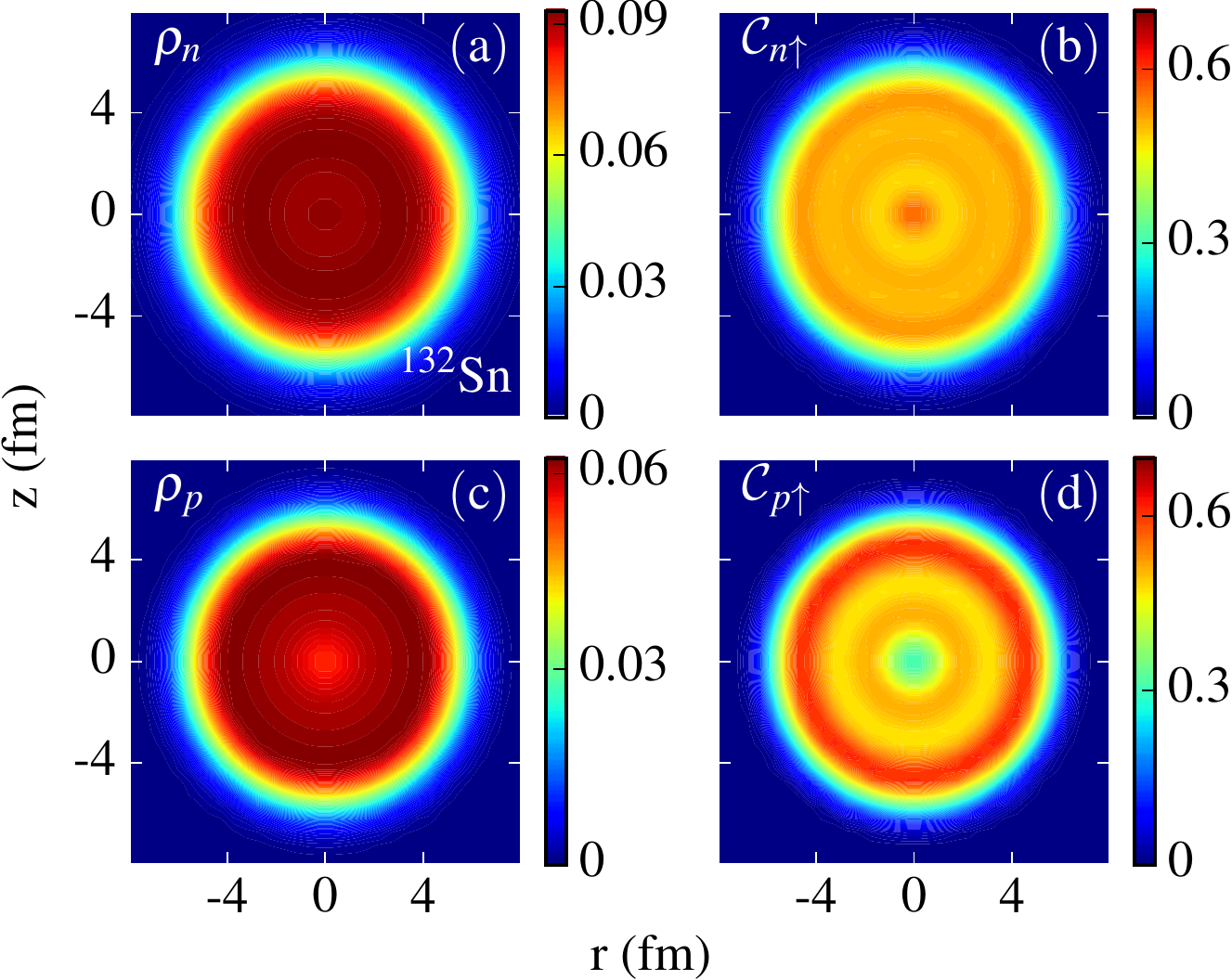} 
   \caption{Nucleonic densities  (in nucleons/fm$^3$) and spatial localizations for the ground state of $^{132}$Sn.}
   \label{fig:sn132}
\end{figure}

Figure~\ref{fig:fm264} shows neutron and proton densities and NLFs for $^{264}$Fm along the fission pathway. We choose three very elongated configurations corresponding to  decreasing  neck sizes. To study the gradual emergence of  fission fragments, we  performed HFB calculations for the ground state densities and NLFs of  $^{132}$Sn, see Fig.~\ref{fig:sn132}. The nucleus $^{132}$Sn is a doubly-magic system with a characteristic shell structure. Except for a small depression at the center of proton density in Fig.~\ref{fig:sn132}(c), the nucleonic  densities are almost constant in the interior. On the other hand, the NLFs show patterns of concentric rings with enhanced localization, in which the neutron NLF exhibits one additional maximum as compared to the proton NLF; this is due to the additional closed neutron shell. As one can see, unlike in atomic systems \cite{Becke1990}, 
the total number of shells cannot be directly read from the number of peaks in the NLF, because the radial distributions of  wave functions belonging to different nucleonic shells vary fairly smoothly and are poorly separated in space. Nevertheless, each magic number leaves a strong and unique imprint on the spatial localization. 
\begin{figure}[htb]
   \includegraphics[width=0.8\linewidth]{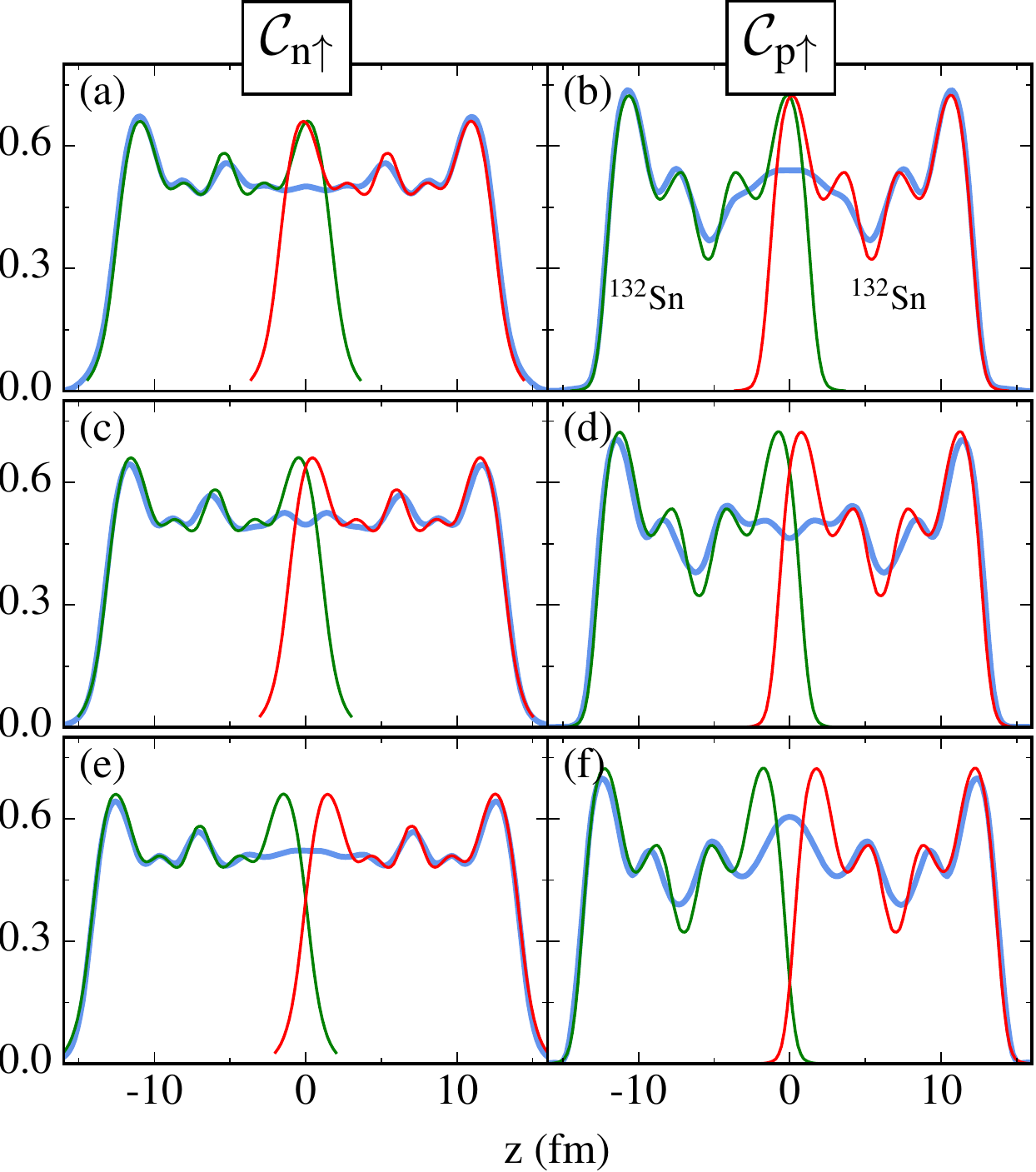} 
   \caption{Neutron (left) and proton (right) NLF profiles for $^{264}$Fm (blue thick line) and two $^{132}$Sn (red and green line) along the $z$ axis ($r=0$). The three panels (a-b), (c-d), and (e-f) correspond to three  deformed configurations of Fig.~\ref{fig:fm264}.}
   \label{fig:loc-line0}
\end{figure}
By comparing the results of Figs.~\ref{fig:fm264} and \ref{fig:sn132} one can clearly see the gradual development of the  $^{132}$Sn clusters within the fissioning $^{264}$Fm. 
It is striking to see that the  ring-like pattern of NLFs develops at an  early stage  of  fission, at which the neck is hardly formed. To illustrate this point   more clearly, Fig.~\ref{fig:loc-line0} displays the NLFs  for the  elongated configurations of $^{264}$Fm shown in Fig.~\ref{fig:fm264} along the $z$-axis and compares them to  those of $^{132}$Sn.  To avoid normalization problems we present  NLFs given by Eq.~(\ref{eqn:localization}), i.e., without applying the density form factor. 
It is seen that the localizations of the emerging fragments match  those of  $^{132}$Sn fairly well in the exterior region.

\begin{figure}[htb]
   \includegraphics[width=\linewidth]{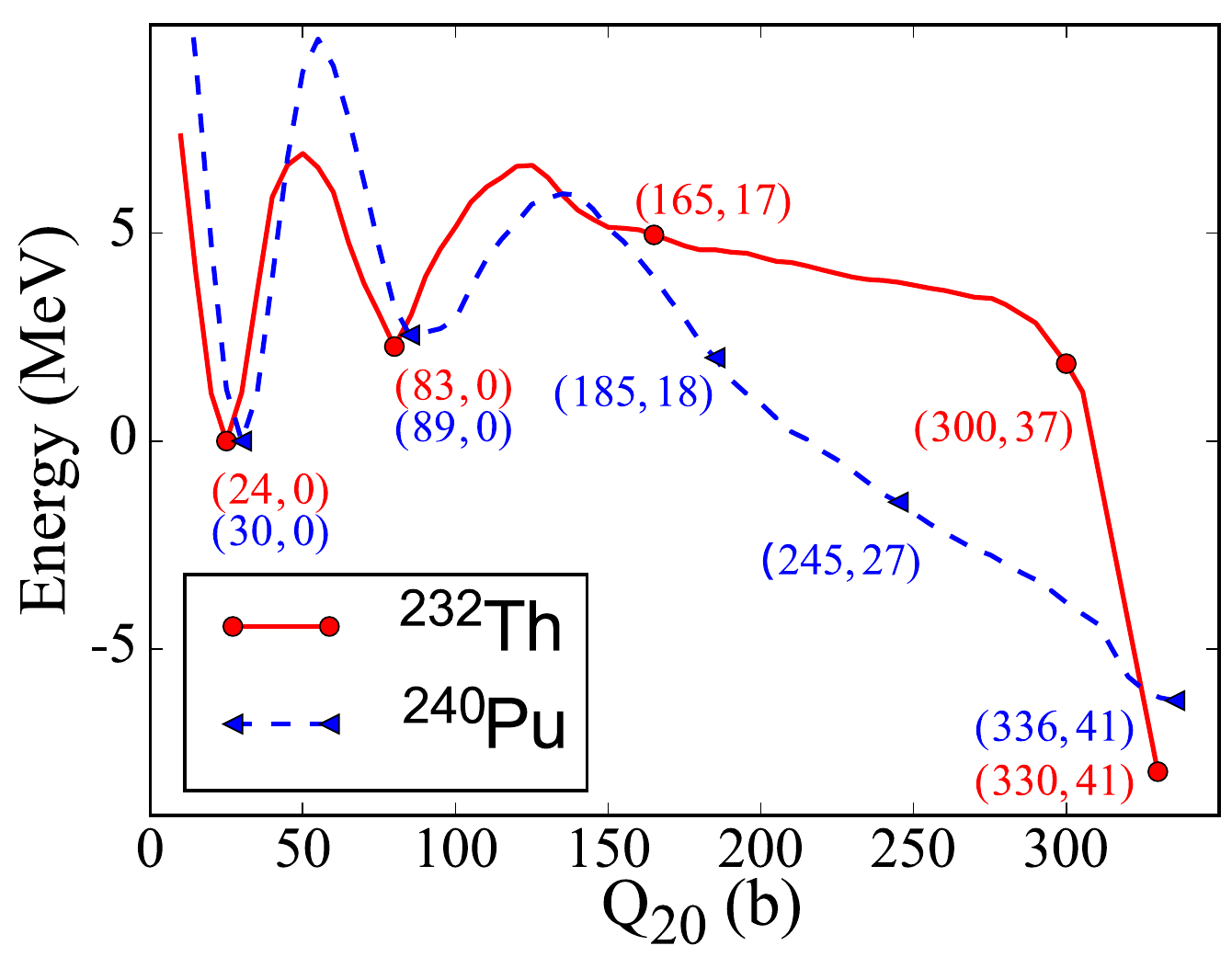} 
   \caption{The potential energy curves of $^{232}$Th and $^{240}$Pu calculated with UNEDF1 along the fission pathways \cite{McDonnell2013,Sadhukhan2016}. The configurations further discussed in Figs.~\ref{fig:th232} and \ref{fig:pu240} are marked by symbols. Their quadrupole and octupole moments, $Q_{20}$(b) and $Q_{30}$ (b$^{3/2}$) respectively,  are indicated.} 
   \label{fig:PES}
\end{figure}

Let us now discuss  two examples of asymmetric fission.
Figure~\ref{fig:PES} shows the potential energy curves of $^{232}$Th and $^{240}$Pu along the most probable fission pathway predicted, respectively, in Refs.~\cite{McDonnell2013} and \cite{Sadhukhan2016}. Both curves show  secondary minima associated with the superdeformed fission isomers. For $^{232}$Th, a pronounced softness is observed at large quadrupole moments  $Q_{20}\approx150-200$\,b. In this region of collective space, a hyperdeformed third minimum is predicted by some Skyrme functionals  \cite{McDonnell2013}. In the next step, we consider five configurations along the fission pathway to perform detailed localization analysis. 

\begin{figure}[htb]
   \includegraphics[width=\linewidth]{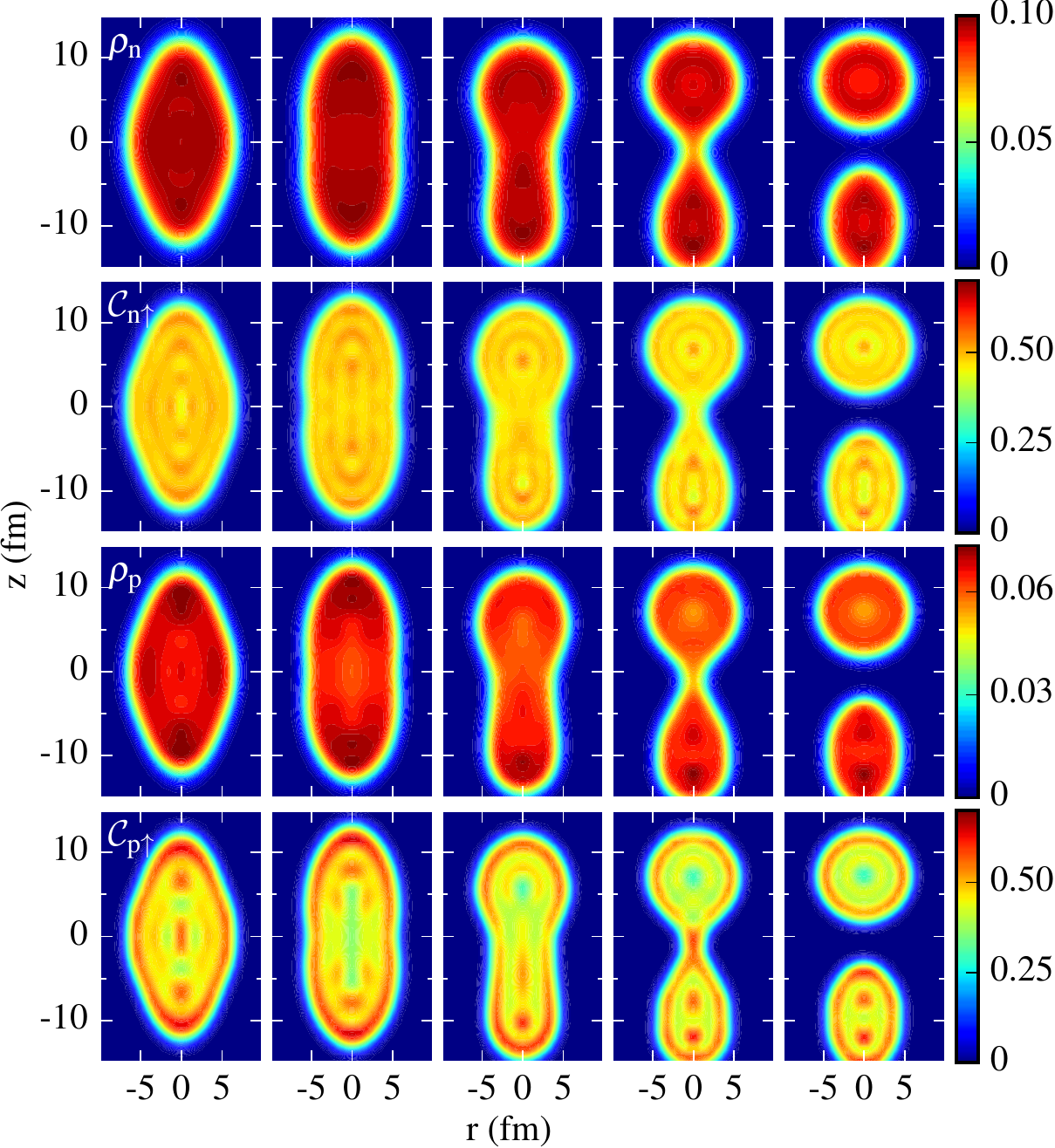} 
   \caption{Nucleonic densities  (in nucleons/fm$^3$) and spatial localizations for $^{232}$Th obtained from HFB calculations with UNEDF1for five configurations along the fission pathway marked in Fig.~\ref{fig:PES}.}
   \label{fig:th232}
\end{figure}

Figure~\ref{fig:th232} shows neutron and proton densities and NLFs for $^{232}$Th along the fission pathway. The first column corresponds to the ground-state configuration where the densities do not show obvious internal structures. However, the neutron NLF shows three concentric ellipses and the proton NLF exhibits two maxima and an enhancement at the surface. The second column corresponds to the fission isomer. Here two-center distributions begin to form in both NLFs.  As discussed in \cite{McDonnell2013}, the distributions  shown in the third column can be  associated with a quasimolecular ``third-minimum" configuration, in which one fragment bears a strong resemblance to the doubly magic nucleus $^{132}$Sn. The forth column represents the configuration close to the scission point, where two well-developed fragments are present. As seen in the last column, at larger elongations the nucleus breaks up into two fragments, one spherical and another one strongly deformed shape.

\begin{figure}[htb]
   \includegraphics[width=\linewidth]{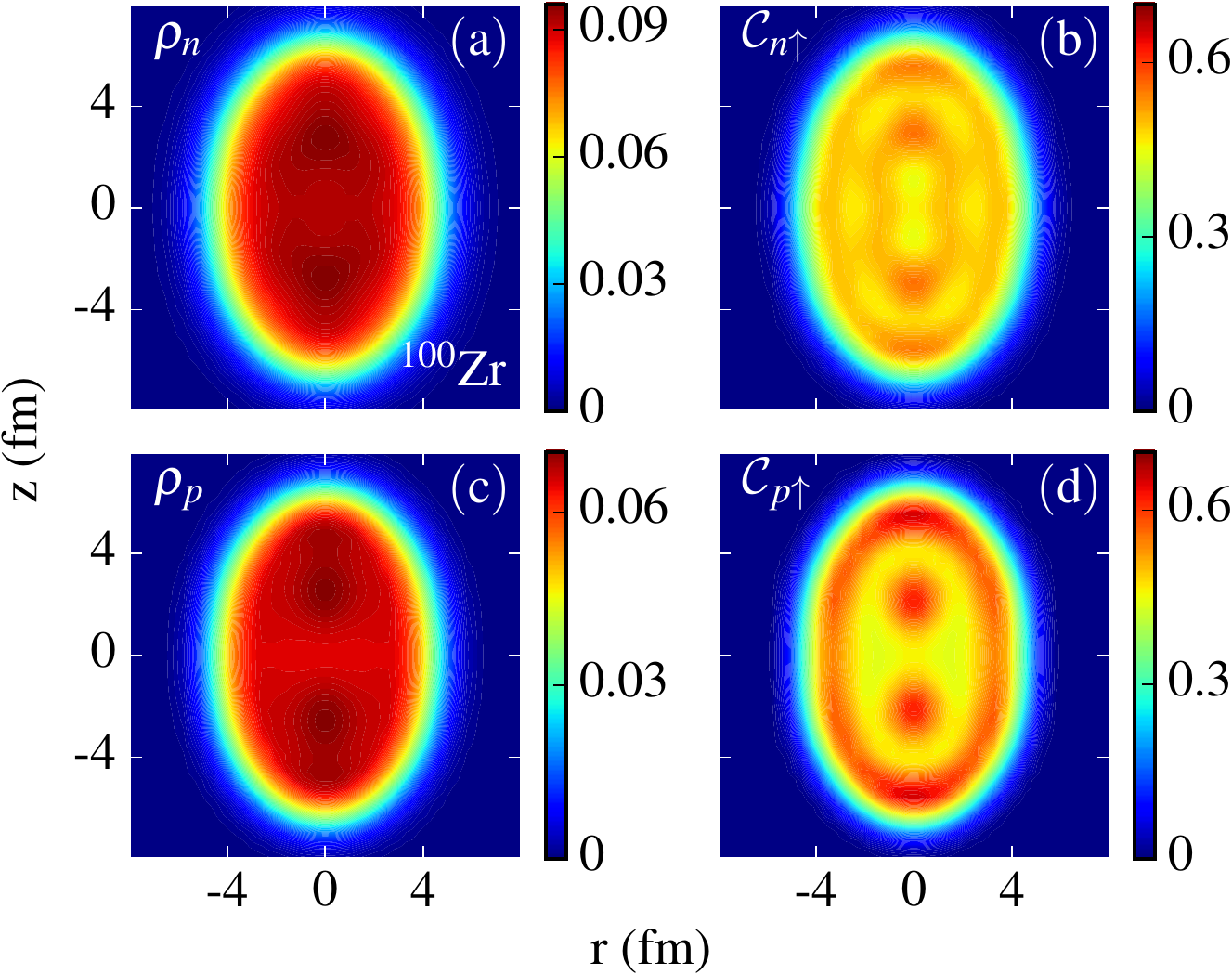} 
   \caption{Similar to Fig.~\ref{fig:sn132}, but for $^{100}$Zr.}
   \label{fig:zr100}
\end{figure}

\begin{figure}[htb]
   \includegraphics[width=0.8\linewidth]{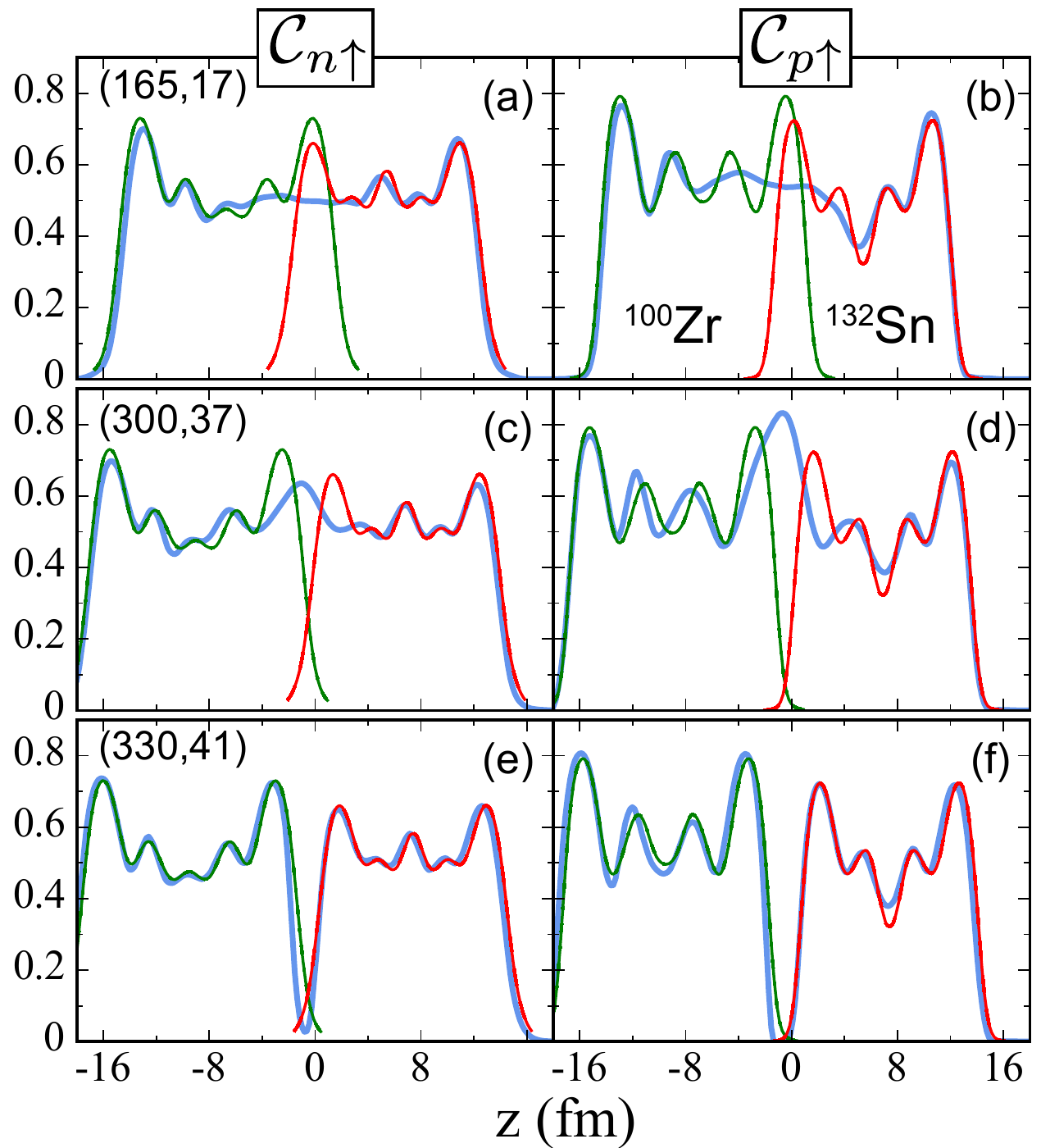} 
   \caption{Neutron (left) and proton (right) NLF profiles for $^{232}$Th (blue thick line), $^{100}$Zr (green line), and $^{132}$Sn (red line)  along the $z$ axis ($r=0$). The first, second, and third panel correspond to the  configurations in the third, fourth and fifth columns of Fig.~\ref{fig:th232}, respectively.}
   \label{fig:loc-line}
\end{figure}

To study the evolution of fission fragments, in addition to $^{132}$Sn  (Fig.~\ref{fig:sn132}) we study  $^{100}$Zr, which is the second presumed fission product of $^{232}$Th. 
The calculation for  $^{100}$Zr is performed at the prolate configuration with ${Q}_{20}=10$\,b, which corresponds to the lighter fission fragment  predicted in \cite{McDonnell2013}. The results are shown in Fig.~\ref{fig:zr100}. Again, while the particle   densities are almost constant in the interior,  the neutron NLF shows two concentric ovals and the proton NLF exhibits two centers in the interior and one enhanced oval at the surface.

The characteristic patterns seen in the NLFs of fission fragments can  be spotted during  the evolution of  $^{232}$Th in Fig.~\ref{fig:th232}. To show it more clearly, Fig.~\ref{fig:loc-line} displays the NLFs  of the  three most elongated configurations of $^{232}$Th along the $z$-axis in Fig.~\ref{fig:th232} and compares them to  those of $^{132}$Sn and $^{100}$Zr.   In Figs.~\ref{fig:loc-line} (a) and (b),  neutron and proton localizations at the center are around 0.5, which is close to the Fermi gas limit. This is expected for a fairly heavy nucleus.
In the exterior, the localizations of two developing fragments match  those of $^{100}$Zr and $^{132}$Sn fairly well. In panels (c) and (d), the NLFs of $^{232}$Th grow in the interior; this  demonstrates that the nucleons become localized at the  neck region.  Finally, in panels (e) and (f), the fission fragments are  separated and their NLFs are consistent with the localizations of  $^{100}$Zr and $^{132}$Sn.

\begin{figure}[htb]
   \includegraphics[width=\linewidth]{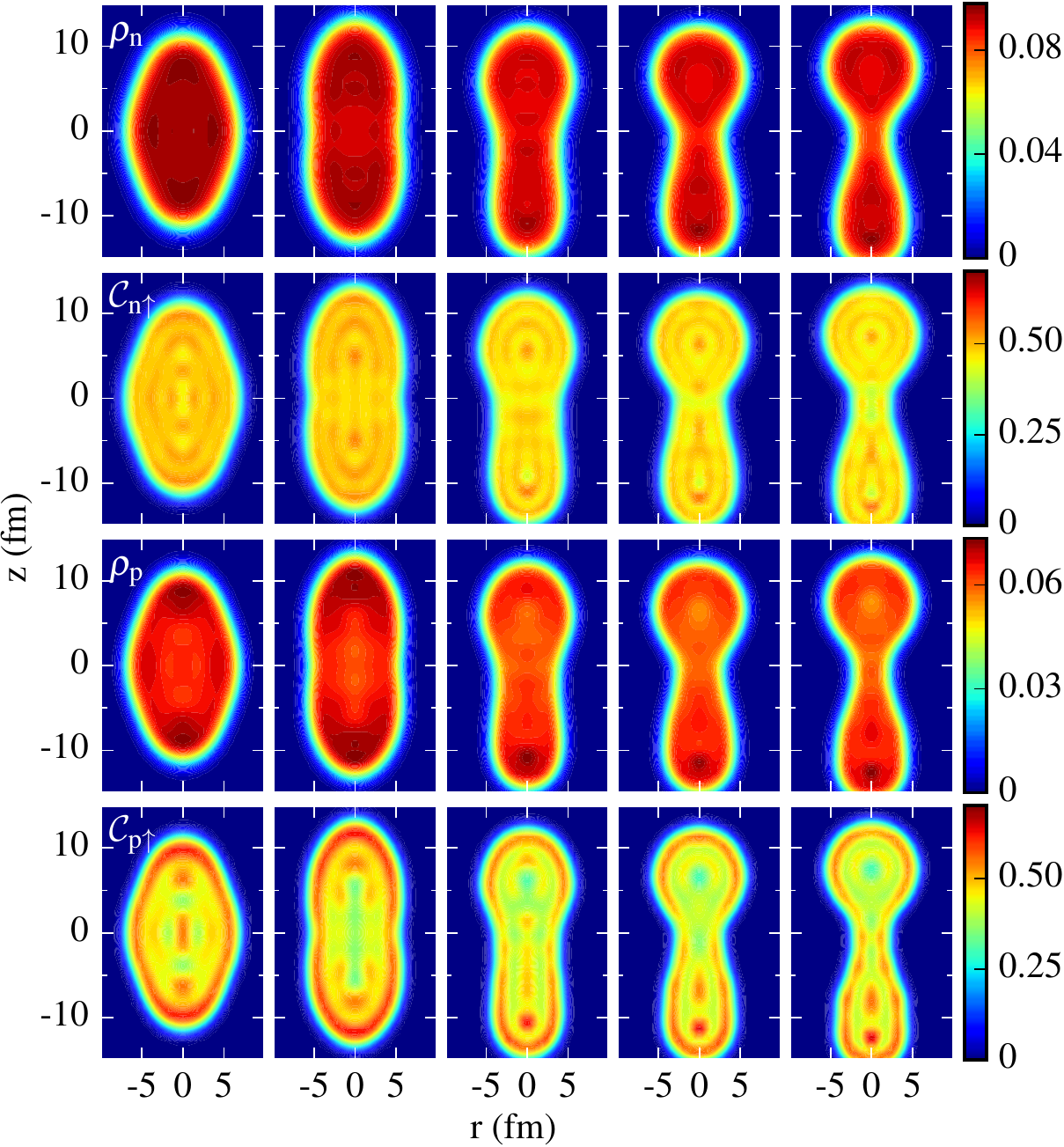} 
   \caption{Similar to Fig.~\ref{fig:th232} but for the configurations of $^{240}$Pu indicated in Fig.~\ref{fig:PES}.}
   \label{fig:pu240}
\end{figure}
Finally, let us consider the important case of $^{240}$Pu. Recently, a microscopic modeling of mass and charge distributions in spontaneous fission of this nucleus was carried out in Ref.~\cite{Sadhukhan2016}. To give an insight into the evolution of $^{240}$Pu along its fission pathway, in Fig.~\ref{fig:pu240} we  illustrate the NLFs of $^{240}$Pu. The transition to the reflection-asymmetric pathway begins  at $Q_{20}\approx 95$\,b. It is seen that  two nascent fragments  start developing  at this configuration. At larger elongations internal parity  is broken and two fragments are formed with distinct shell imprints in the corresponding NLFs. In the last column, the rings of enhanced localization are almost closed, and the fragments are nearly separated.

The examples presented above  show in a rather dramatic fashion that the NLFs can serve as  excellent fingerprints of both the formation and evolution of  cluster structures in fissioning nuclei.

\section{Conclusion}
\label{summary}

In this work, we presented  DFT developments pertaining to the theoretical description of fragment emergence in heavy  fissioning systems. Building upon results of previous work on cluster formation
in light nuclei \cite{Reinhard2011}, we demonstrated that the nucleon  spatial localization is a superb indicator of clustering in heavy nuclei; the  characteristic patterns of NLFs can serve as fingerprints of the single-particle shell structure associated with cluster configurations. 

While the characteristic oscillating pattern of  NLFs magnifies the cluster structures in light nuclei \cite{Reinhard2011}, shell effects in nascent fragments in   fissioning nuclei also  leave a strong imprint on the localization. Our EDFM analysis of fission evolution of $^{264}$Fm, $^{232}$Th, and $^{240}$Pu demonstrates  that the fragments are formed fairly early in the evolution, well before scission.

Future applications of NLFs will include studies of clustering in medium-mass nuclei as well as the identification of fission yields. Another interesting use of NLFs will be in the description and visualization of collective rotational motion, where spin-up and spin-down localizations are different due to 
the breaking of time-reversal symmetry. Such a study will provide insights into the angular momentum dynamics in atomic nuclei.

\begin{acknowledgements}
Useful discussions with P.-G. Reinhard, N. Schunck, and A.S. Umar, and the help of E. Olsen are gratefully acknowledged. This work was supported by the U.S. Department of Energy, Office of Science under Award Numbers DOE-DE-NA0002847 (the Stewardship Science Academic Alliances program), DE-SC0013365 (Michigan State University), and DE-SC0008511 (NUCLEI SciDAC-3 collaboration). An award of computer time was provided by the Institute for Cyber-Enabled Research at Michigan State University. We also used computational resource provided by the National Energy Research Scientific Computing Center (NERSC).
\end{acknowledgements}
%

\begin{thebibliography}{56}%
\makeatletter
\providecommand \@ifxundefined [1]{%
 \@ifx{#1\undefined}
}%
\providecommand \@ifnum [1]{%
 \ifnum #1\expandafter \@firstoftwo
 \else \expandafter \@secondoftwo
 \fi
}%
\providecommand \@ifx [1]{%
 \ifx #1\expandafter \@firstoftwo
 \else \expandafter \@secondoftwo
 \fi
}%
\providecommand \natexlab [1]{#1}%
\providecommand \enquote  [1]{``#1''}%
\providecommand \bibnamefont  [1]{#1}%
\providecommand \bibfnamefont [1]{#1}%
\providecommand \citenamefont [1]{#1}%
\providecommand \href@noop [0]{\@secondoftwo}%
\providecommand \href [0]{\begingroup \@sanitize@url \@href}%
\providecommand \@href[1]{\@@startlink{#1}\@@href}%
\providecommand \@@href[1]{\endgroup#1\@@endlink}%
\providecommand \@sanitize@url [0]{\catcode `\\12\catcode `\$12\catcode
  `\&12\catcode `\#12\catcode `\^12\catcode `\_12\catcode `\%12\relax}%
\providecommand \@@startlink[1]{}%
\providecommand \@@endlink[0]{}%
\providecommand \url  [0]{\begingroup\@sanitize@url \@url }%
\providecommand \@url [1]{\endgroup\@href {#1}{\urlprefix }}%
\providecommand \urlprefix  [0]{URL }%
\providecommand \Eprint [0]{\href }%
\providecommand \doibase [0]{http://dx.doi.org/}%
\providecommand \selectlanguage [0]{\@gobble}%
\providecommand \bibinfo  [0]{\@secondoftwo}%
\providecommand \bibfield  [0]{\@secondoftwo}%
\providecommand \translation [1]{[#1]}%
\providecommand \BibitemOpen [0]{}%
\providecommand \bibitemStop [0]{}%
\providecommand \bibitemNoStop [0]{.\EOS\space}%
\providecommand \EOS [0]{\spacefactor3000\relax}%
\providecommand \BibitemShut  [1]{\csname bibitem#1\endcsname}%
\let\auto@bib@innerbib\@empty
\bibitem [{\citenamefont {Beck}(2010)}]{beck2010clusters}%
  \BibitemOpen
  \bibfield  {author} {\bibinfo {author} {\bibfnamefont {C.}~\bibnamefont
  {Beck}},\ }\href@noop {} {\emph {\bibinfo {title} {Clusters in Nuclei}}},\
  Vol.~\bibinfo {volume} {1}\ (\bibinfo  {publisher} {Springer-Verlag Berlin
  Heidelberg},\ \bibinfo {year} {2010})\BibitemShut {NoStop}%
\bibitem [{\citenamefont {Beck}(2012)}]{beck2012clusters}%
  \BibitemOpen
  \bibfield  {author} {\bibinfo {author} {\bibfnamefont {C.}~\bibnamefont
  {Beck}},\ }\href@noop {} {\emph {\bibinfo {title} {Clusters in Nuclei}}},\
  Vol.~\bibinfo {volume} {2}\ (\bibinfo  {publisher} {Springer-Verlag Berlin
  Heidelberg},\ \bibinfo {year} {2012})\BibitemShut {NoStop}%
\bibitem [{\citenamefont {Beck}(2014)}]{beck2014clusters}%
  \BibitemOpen
  \bibfield  {author} {\bibinfo {author} {\bibfnamefont {C.}~\bibnamefont
  {Beck}},\ }\href@noop {} {\emph {\bibinfo {title} {Clusters in Nuclei}}},\
  Vol.~\bibinfo {volume} {3}\ (\bibinfo  {publisher} {Springer International
  Publishing},\ \bibinfo {year} {2014})\BibitemShut {NoStop}%
\bibitem [{\citenamefont {von Oertzen}\ \emph {et~al.}(2006)\citenamefont {von
  Oertzen}, \citenamefont {Freer},\ and\ \citenamefont
  {Kanada-En$^{'}$yo}}]{vonOertzen2006}%
  \BibitemOpen
  \bibfield  {author} {\bibinfo {author} {\bibfnamefont {W.}~\bibnamefont {von
  Oertzen}}, \bibinfo {author} {\bibfnamefont {M.}~\bibnamefont {Freer}}, \
  and\ \bibinfo {author} {\bibfnamefont {Y.}~\bibnamefont
  {Kanada-En$^{'}$yo}},\ }\href {\doibase
  http://dx.doi.org/10.1016/j.physrep.2006.07.001} {\bibfield  {journal}
  {\bibinfo  {journal} {Phys. Rep.}\ }\textbf {\bibinfo {volume} {432}},\
  \bibinfo {pages} {43 } (\bibinfo {year} {2006})}\BibitemShut {NoStop}%
\bibitem [{\citenamefont {Delion}(2010)}]{DelionCluster}%
  \BibitemOpen
  \bibfield  {author} {\bibinfo {author} {\bibfnamefont {D.~S.}\ \bibnamefont
  {Delion}},\ }\href@noop {} {\emph {\bibinfo {title} {Theory of Particle and
  Cluster Emission}}},\ \bibinfo {series} {Lecture Notes in Physics}, Vol.\
  \bibinfo {volume} {829}\ (\bibinfo  {publisher} {Springer-Verlag Berlin
  Heidelberg},\ \bibinfo {year} {2010})\BibitemShut {NoStop}%
\bibitem [{\citenamefont {Hafstad}\ and\ \citenamefont
  {Teller}(1938)}]{Hafstad}%
  \BibitemOpen
  \bibfield  {author} {\bibinfo {author} {\bibfnamefont {L.~R.}\ \bibnamefont
  {Hafstad}}\ and\ \bibinfo {author} {\bibfnamefont {E.}~\bibnamefont
  {Teller}},\ }\href {\doibase 10.1103/PhysRev.54.681} {\bibfield  {journal}
  {\bibinfo  {journal} {Phys. Rev.}\ }\textbf {\bibinfo {volume} {54}},\
  \bibinfo {pages} {681} (\bibinfo {year} {1938})}\BibitemShut {NoStop}%
\bibitem [{\citenamefont {Rose}\ and\ \citenamefont {Jones}(1984)}]{Rose1984}%
  \BibitemOpen
  \bibfield  {author} {\bibinfo {author} {\bibfnamefont {H.~J.}\ \bibnamefont
  {Rose}}\ and\ \bibinfo {author} {\bibfnamefont {G.~A.}\ \bibnamefont
  {Jones}},\ }\href {http://dx.doi.org/10.1038/307245a0} {\bibfield  {journal}
  {\bibinfo  {journal} {Nature}\ }\textbf {\bibinfo {volume} {307}},\ \bibinfo
  {pages} {245} (\bibinfo {year} {1984})}\BibitemShut {NoStop}%
\bibitem [{\citenamefont {{Aleksandrov}}\ \emph {et~al.}(1984)\citenamefont
  {{Aleksandrov}}, \citenamefont {{Belyatski{\v i}}}, \citenamefont
  {{Glukhov}}, \citenamefont {{Nikol'Ski{\v i}}}, \citenamefont {{Novatski{\v
  i}}}, \citenamefont {{Ogloblin}},\ and\ \citenamefont
  {{Stepanov}}}]{Aleksandrov1984}%
  \BibitemOpen
  \bibfield  {author} {\bibinfo {author} {\bibfnamefont {D.~V.}\ \bibnamefont
  {{Aleksandrov}}}, \bibinfo {author} {\bibfnamefont {A.~F.}\ \bibnamefont
  {{Belyatski{\v i}}}}, \bibinfo {author} {\bibfnamefont {Y.~A.}\ \bibnamefont
  {{Glukhov}}}, \bibinfo {author} {\bibfnamefont {E.~Y.}\ \bibnamefont
  {{Nikol'Ski{\v i}}}}, \bibinfo {author} {\bibfnamefont {B.~G.}\ \bibnamefont
  {{Novatski{\v i}}}}, \bibinfo {author} {\bibfnamefont {A.~A.}\ \bibnamefont
  {{Ogloblin}}}, \ and\ \bibinfo {author} {\bibfnamefont {D.~N.}\ \bibnamefont
  {{Stepanov}}},\ }\href@noop {} {\bibfield  {journal} {\bibinfo  {journal}
  {ZhETF Pisma Redaktsiiu}\ }\textbf {\bibinfo {volume} {40}},\ \bibinfo
  {pages} {152} (\bibinfo {year} {1984})}\BibitemShut {NoStop}%
\bibitem [{\citenamefont {Bj\o{}rnholm}\ and\ \citenamefont
  {Lynn}(1980)}]{BLynn}%
  \BibitemOpen
  \bibfield  {author} {\bibinfo {author} {\bibfnamefont {S.}~\bibnamefont
  {Bj\o{}rnholm}}\ and\ \bibinfo {author} {\bibfnamefont {J.~E.}\ \bibnamefont
  {Lynn}},\ }\href {\doibase 10.1103/RevModPhys.52.725} {\bibfield  {journal}
  {\bibinfo  {journal} {Rev. Mod. Phys.}\ }\textbf {\bibinfo {volume} {52}},\
  \bibinfo {pages} {725} (\bibinfo {year} {1980})}\BibitemShut {NoStop}%
\bibitem [{\citenamefont {R{\"o}pke}\ \emph {et~al.}(2006)\citenamefont
  {R{\"o}pke}, \citenamefont {Grigo}, \citenamefont {Sumiyoshi},\ and\
  \citenamefont {Shen}}]{Ropke2006}%
  \BibitemOpen
  \bibfield  {author} {\bibinfo {author} {\bibfnamefont {G.}~\bibnamefont
  {R{\"o}pke}}, \bibinfo {author} {\bibfnamefont {A.}~\bibnamefont {Grigo}},
  \bibinfo {author} {\bibfnamefont {K.}~\bibnamefont {Sumiyoshi}}, \ and\
  \bibinfo {author} {\bibfnamefont {H.}~\bibnamefont {Shen}},\ }\enquote
  {\bibinfo {title} {Clusters and condensates in the nuclear matter equation of
  state},}\ in\ \href {\doibase 10.1007/1-4020-3430-X_05} {\emph {\bibinfo
  {booktitle} {Superdense QCD Matter and Compact Stars}}},\ \bibinfo {editor}
  {edited by\ \bibinfo {editor} {\bibfnamefont {D.}~\bibnamefont {Blaschke}}\
  and\ \bibinfo {editor} {\bibfnamefont {D.}~\bibnamefont {Sedrakian}}}\
  (\bibinfo  {publisher} {Springer Netherlands},\ \bibinfo {address}
  {Dordrecht},\ \bibinfo {year} {2006})\ p.~\bibinfo {pages} {73}\BibitemShut
  {NoStop}%
\bibitem [{\citenamefont {Girod}\ and\ \citenamefont {Schuck}(2013)}]{Girod13}%
  \BibitemOpen
  \bibfield  {author} {\bibinfo {author} {\bibfnamefont {M.}~\bibnamefont
  {Girod}}\ and\ \bibinfo {author} {\bibfnamefont {P.}~\bibnamefont {Schuck}},\
  }\href {\doibase 10.1103/PhysRevLett.111.132503} {\bibfield  {journal}
  {\bibinfo  {journal} {Phys. Rev. Lett.}\ }\textbf {\bibinfo {volume} {111}},\
  \bibinfo {pages} {132503} (\bibinfo {year} {2013})}\BibitemShut {NoStop}%
\bibitem [{\citenamefont {Ebran}\ \emph
  {et~al.}(2014{\natexlab{a}})\citenamefont {Ebran}, \citenamefont {Khan},
  \citenamefont {Nik\ifmmode \check{s}\else \v{s}\fi{}i\ifmmode~\acute{c}\else
  \'{c}\fi{}},\ and\ \citenamefont {Vretenar}}]{Ebran14a}%
  \BibitemOpen
  \bibfield  {author} {\bibinfo {author} {\bibfnamefont {J.-P.}\ \bibnamefont
  {Ebran}}, \bibinfo {author} {\bibfnamefont {E.}~\bibnamefont {Khan}},
  \bibinfo {author} {\bibfnamefont {T.}~\bibnamefont {Nik\ifmmode
  \check{s}\else \v{s}\fi{}i\ifmmode~\acute{c}\else \'{c}\fi{}}}, \ and\
  \bibinfo {author} {\bibfnamefont {D.}~\bibnamefont {Vretenar}},\ }\href
  {\doibase 10.1103/PhysRevC.89.031303} {\bibfield  {journal} {\bibinfo
  {journal} {Phys. Rev. C}\ }\textbf {\bibinfo {volume} {89}},\ \bibinfo
  {pages} {031303} (\bibinfo {year} {2014}{\natexlab{a}})}\BibitemShut
  {NoStop}%
\bibitem [{\citenamefont {Brink}\ and\ \citenamefont
  {Castro}(1973)}]{BRINK1973109}%
  \BibitemOpen
  \bibfield  {author} {\bibinfo {author} {\bibfnamefont {D.}~\bibnamefont
  {Brink}}\ and\ \bibinfo {author} {\bibfnamefont {J.}~\bibnamefont {Castro}},\
  }\href {\doibase http://dx.doi.org/10.1016/0375-9474(73)90521-6} {\bibfield
  {journal} {\bibinfo  {journal} {Nucl. Phys. A}\ }\textbf {\bibinfo {volume}
  {216}},\ \bibinfo {pages} {109} (\bibinfo {year} {1973})}\BibitemShut
  {NoStop}%
\bibitem [{\citenamefont {Oko{\l}owicz}\ \emph {et~al.}(2012)\citenamefont
  {Oko{\l}owicz}, \citenamefont {P{\l}oszajczak},\ and\ \citenamefont
  {Nazarewicz}}]{Okolowicz12}%
  \BibitemOpen
  \bibfield  {author} {\bibinfo {author} {\bibfnamefont {J.}~\bibnamefont
  {Oko{\l}owicz}}, \bibinfo {author} {\bibfnamefont {M.}~\bibnamefont
  {P{\l}oszajczak}}, \ and\ \bibinfo {author} {\bibfnamefont {W.}~\bibnamefont
  {Nazarewicz}},\ }\href {\doibase 10.1143/PTPS.196.230} {\bibfield  {journal}
  {\bibinfo  {journal} {Prog. Theor. Phys. Suppl.}\ }\textbf {\bibinfo {volume}
  {196}},\ \bibinfo {pages} {230} (\bibinfo {year} {2012})}\BibitemShut
  {NoStop}%
\bibitem [{\citenamefont {Oko{\l}owicz}\ \emph {et~al.}(2013)\citenamefont
  {Oko{\l}owicz}, \citenamefont {Nazarewicz},\ and\ \citenamefont
  {P{\l}oszajczak}}]{Okolowicz13}%
  \BibitemOpen
  \bibfield  {author} {\bibinfo {author} {\bibfnamefont {J.}~\bibnamefont
  {Oko{\l}owicz}}, \bibinfo {author} {\bibfnamefont {W.}~\bibnamefont
  {Nazarewicz}}, \ and\ \bibinfo {author} {\bibfnamefont {M.}~\bibnamefont
  {P{\l}oszajczak}},\ }\href {\doibase 10.1002/prop.201200127} {\bibfield
  {journal} {\bibinfo  {journal} {Fortschr. Phys.}\ }\textbf {\bibinfo {volume}
  {61}},\ \bibinfo {pages} {66} (\bibinfo {year} {2013})}\BibitemShut {NoStop}%
\bibitem [{\citenamefont {Ikeda}\ \emph {et~al.}(1968)\citenamefont {Ikeda},
  \citenamefont {Takigawa},\ and\ \citenamefont {Horiuchi}}]{Ikeda68}%
  \BibitemOpen
  \bibfield  {author} {\bibinfo {author} {\bibfnamefont {K.}~\bibnamefont
  {Ikeda}}, \bibinfo {author} {\bibfnamefont {N.}~\bibnamefont {Takigawa}}, \
  and\ \bibinfo {author} {\bibfnamefont {H.}~\bibnamefont {Horiuchi}},\ }\href
  {\doibase 10.1143/PTPS.E68.464} {\bibfield  {journal} {\bibinfo  {journal}
  {Prog. Theor. Phys. Suppl.}\ }\textbf {\bibinfo {volume} {E68}},\ \bibinfo
  {pages} {464} (\bibinfo {year} {1968})}\BibitemShut {NoStop}%
\bibitem [{\citenamefont {Lovas}\ \emph {et~al.}(2013)\citenamefont {Lovas},
  \citenamefont {Dombr{\'a}di}, \citenamefont {Kiss}, \citenamefont {Kruppa},\
  and\ \citenamefont {L{\'e}vai}}]{Cluster12}%
  \BibitemOpen
  \bibinfo {editor} {\bibfnamefont {R.~G.}\ \bibnamefont {Lovas}}, \bibinfo
  {editor} {\bibfnamefont {Z.}~\bibnamefont {Dombr{\'a}di}}, \bibinfo {editor}
  {\bibfnamefont {G.~G.}\ \bibnamefont {Kiss}}, \bibinfo {editor}
  {\bibfnamefont {A.~T.}\ \bibnamefont {Kruppa}}, \ and\ \bibinfo {editor}
  {\bibfnamefont {G.}~\bibnamefont {L{\'e}vai}},\ eds.,\ \href@noop {} {\emph
  {\bibinfo {title} {10th International Conference on Clustering Aspects of
  Nuclear Structure and Dynamics, J. Phys. Conf. Ser.}}},\ Vol.\ \bibinfo
  {volume} {436}\ (\bibinfo {year} {2013})\BibitemShut {NoStop}%
\bibitem [{\citenamefont {Yoshida}\ \emph {et~al.}(2013)\citenamefont
  {Yoshida}, \citenamefont {Shimizu}, \citenamefont {Abe},\ and\ \citenamefont
  {Otsuka}}]{Yoshida13}%
  \BibitemOpen
  \bibfield  {author} {\bibinfo {author} {\bibfnamefont {T.}~\bibnamefont
  {Yoshida}}, \bibinfo {author} {\bibfnamefont {N.}~\bibnamefont {Shimizu}},
  \bibinfo {author} {\bibfnamefont {T.}~\bibnamefont {Abe}}, \ and\ \bibinfo
  {author} {\bibfnamefont {T.}~\bibnamefont {Otsuka}},\ }\href
  {http://stacks.iop.org/1742-6596/454/i=1/a=012050} {\bibfield  {journal}
  {\bibinfo  {journal} {J. Phys. Conf. Ser.}\ }\textbf {\bibinfo {volume}
  {454}},\ \bibinfo {pages} {012050} (\bibinfo {year} {2013})}\BibitemShut
  {NoStop}%
\bibitem [{\citenamefont {Epelbaum}\ \emph {et~al.}(2012)\citenamefont
  {Epelbaum}, \citenamefont {Krebs}, \citenamefont {L\"ahde}, \citenamefont
  {Lee},\ and\ \citenamefont {Mei\ss{}ner}}]{Epelbaum12}%
  \BibitemOpen
  \bibfield  {author} {\bibinfo {author} {\bibfnamefont {E.}~\bibnamefont
  {Epelbaum}}, \bibinfo {author} {\bibfnamefont {H.}~\bibnamefont {Krebs}},
  \bibinfo {author} {\bibfnamefont {T.~A.}\ \bibnamefont {L\"ahde}}, \bibinfo
  {author} {\bibfnamefont {D.}~\bibnamefont {Lee}}, \ and\ \bibinfo {author}
  {\bibfnamefont {U.-G.}\ \bibnamefont {Mei\ss{}ner}},\ }\href {\doibase
  10.1103/PhysRevLett.109.252501} {\bibfield  {journal} {\bibinfo  {journal}
  {Phys. Rev. Lett.}\ }\textbf {\bibinfo {volume} {109}},\ \bibinfo {pages}
  {252501} (\bibinfo {year} {2012})}\BibitemShut {NoStop}%
\bibitem [{\citenamefont {Epelbaum}\ \emph {et~al.}(2014)\citenamefont
  {Epelbaum}, \citenamefont {Krebs}, \citenamefont {L\"ahde}, \citenamefont
  {Lee}, \citenamefont {Mei\ss{}ner},\ and\ \citenamefont
  {Rupak}}]{Epelbaum14}%
  \BibitemOpen
  \bibfield  {author} {\bibinfo {author} {\bibfnamefont {E.}~\bibnamefont
  {Epelbaum}}, \bibinfo {author} {\bibfnamefont {H.}~\bibnamefont {Krebs}},
  \bibinfo {author} {\bibfnamefont {T.~A.}\ \bibnamefont {L\"ahde}}, \bibinfo
  {author} {\bibfnamefont {D.}~\bibnamefont {Lee}}, \bibinfo {author}
  {\bibfnamefont {U.-G.}\ \bibnamefont {Mei\ss{}ner}}, \ and\ \bibinfo {author}
  {\bibfnamefont {G.}~\bibnamefont {Rupak}},\ }\href {\doibase
  10.1103/PhysRevLett.112.102501} {\bibfield  {journal} {\bibinfo  {journal}
  {Phys. Rev. Lett.}\ }\textbf {\bibinfo {volume} {112}},\ \bibinfo {pages}
  {102501} (\bibinfo {year} {2014})}\BibitemShut {NoStop}%
\bibitem [{\citenamefont {Elhatisari}\ \emph {et~al.}(2015)\citenamefont
  {Elhatisari}, \citenamefont {Lee}, \citenamefont {Rupak}, \citenamefont
  {Epelbaum}, \citenamefont {Krebs}, \citenamefont {L{\"a}hde}, \citenamefont
  {Luu},\ and\ \citenamefont {Mei{\ss}ner}}]{Elh15}%
  \BibitemOpen
  \bibfield  {author} {\bibinfo {author} {\bibfnamefont {S.}~\bibnamefont
  {Elhatisari}}, \bibinfo {author} {\bibfnamefont {D.}~\bibnamefont {Lee}},
  \bibinfo {author} {\bibfnamefont {G.}~\bibnamefont {Rupak}}, \bibinfo
  {author} {\bibfnamefont {E.}~\bibnamefont {Epelbaum}}, \bibinfo {author}
  {\bibfnamefont {H.}~\bibnamefont {Krebs}}, \bibinfo {author} {\bibfnamefont
  {T.~A.}\ \bibnamefont {L{\"a}hde}}, \bibinfo {author} {\bibfnamefont
  {T.}~\bibnamefont {Luu}}, \ and\ \bibinfo {author} {\bibfnamefont {U.-G.}\
  \bibnamefont {Mei{\ss}ner}},\ }\href {http://dx.doi.org/10.1038/nature16067}
  {\bibfield  {journal} {\bibinfo  {journal} {Nature}\ }\textbf {\bibinfo
  {volume} {528}},\ \bibinfo {pages} {111} (\bibinfo {year}
  {2015})}\BibitemShut {NoStop}%
\bibitem [{\citenamefont {Carlson}\ \emph {et~al.}(2015)\citenamefont
  {Carlson}, \citenamefont {Gandolfi}, \citenamefont {Pederiva}, \citenamefont
  {Pieper}, \citenamefont {Schiavilla}, \citenamefont {Schmidt},\ and\
  \citenamefont {Wiringa}}]{QMC}%
  \BibitemOpen
  \bibfield  {author} {\bibinfo {author} {\bibfnamefont {J.}~\bibnamefont
  {Carlson}}, \bibinfo {author} {\bibfnamefont {S.}~\bibnamefont {Gandolfi}},
  \bibinfo {author} {\bibfnamefont {F.}~\bibnamefont {Pederiva}}, \bibinfo
  {author} {\bibfnamefont {S.~C.}\ \bibnamefont {Pieper}}, \bibinfo {author}
  {\bibfnamefont {R.}~\bibnamefont {Schiavilla}}, \bibinfo {author}
  {\bibfnamefont {K.~E.}\ \bibnamefont {Schmidt}}, \ and\ \bibinfo {author}
  {\bibfnamefont {R.~B.}\ \bibnamefont {Wiringa}},\ }\href {\doibase
  10.1103/RevModPhys.87.1067} {\bibfield  {journal} {\bibinfo  {journal} {Rev.
  Mod. Phys.}\ }\textbf {\bibinfo {volume} {87}},\ \bibinfo {pages} {1067}
  (\bibinfo {year} {2015})}\BibitemShut {NoStop}%
\bibitem [{\citenamefont {Jones}(2015)}]{Jones15}%
  \BibitemOpen
  \bibfield  {author} {\bibinfo {author} {\bibfnamefont {R.~O.}\ \bibnamefont
  {Jones}},\ }\href {\doibase 10.1103/RevModPhys.87.897} {\bibfield  {journal}
  {\bibinfo  {journal} {Rev. Mod. Phys.}\ }\textbf {\bibinfo {volume} {87}},\
  \bibinfo {pages} {897} (\bibinfo {year} {2015})}\BibitemShut {NoStop}%
\bibitem [{\citenamefont {Bender}\ \emph {et~al.}(2003)\citenamefont {Bender},
  \citenamefont {Heenen},\ and\ \citenamefont {Reinhard}}]{bender2003self}%
  \BibitemOpen
  \bibfield  {author} {\bibinfo {author} {\bibfnamefont {M.}~\bibnamefont
  {Bender}}, \bibinfo {author} {\bibfnamefont {P.-H.}\ \bibnamefont {Heenen}},
  \ and\ \bibinfo {author} {\bibfnamefont {P.-G.}\ \bibnamefont {Reinhard}},\
  }\href@noop {} {\bibfield  {journal} {\bibinfo  {journal} {Rev. Mod. Phys.}\
  }\textbf {\bibinfo {volume} {75}},\ \bibinfo {pages} {121} (\bibinfo {year}
  {2003})}\BibitemShut {NoStop}%
\bibitem [{\citenamefont {Leander}\ and\ \citenamefont
  {Larsson}(1975)}]{Leander}%
  \BibitemOpen
  \bibfield  {author} {\bibinfo {author} {\bibfnamefont {G.}~\bibnamefont
  {Leander}}\ and\ \bibinfo {author} {\bibfnamefont {S.}~\bibnamefont
  {Larsson}},\ }\href {\doibase http://dx.doi.org/10.1016/0375-9474(75)91136-7}
  {\bibfield  {journal} {\bibinfo  {journal} {Nucl. Phys. A}\ }\textbf
  {\bibinfo {volume} {239}},\ \bibinfo {pages} {93} (\bibinfo {year}
  {1975})}\BibitemShut {NoStop}%
\bibitem [{\citenamefont {Flocard}\ \emph {et~al.}(1984)\citenamefont
  {Flocard}, \citenamefont {Heenen}, \citenamefont {Krieger},\ and\
  \citenamefont {Weiss}}]{Flocard84}%
  \BibitemOpen
  \bibfield  {author} {\bibinfo {author} {\bibfnamefont {H.}~\bibnamefont
  {Flocard}}, \bibinfo {author} {\bibfnamefont {P.~H.}\ \bibnamefont {Heenen}},
  \bibinfo {author} {\bibfnamefont {S.~J.}\ \bibnamefont {Krieger}}, \ and\
  \bibinfo {author} {\bibfnamefont {M.~S.}\ \bibnamefont {Weiss}},\ }\href
  {\doibase 10.1143/PTP.72.1000} {\bibfield  {journal} {\bibinfo  {journal}
  {Prog. Theor. Phys.}\ }\textbf {\bibinfo {volume} {72}},\ \bibinfo {pages}
  {1000} (\bibinfo {year} {1984})}\BibitemShut {NoStop}%
\bibitem [{\citenamefont {Marsh}\ and\ \citenamefont {Rae}(1986)}]{Marsh86}%
  \BibitemOpen
  \bibfield  {author} {\bibinfo {author} {\bibfnamefont {S.}~\bibnamefont
  {Marsh}}\ and\ \bibinfo {author} {\bibfnamefont {W.}~\bibnamefont {Rae}},\
  }\href {\doibase http://dx.doi.org/10.1016/0370-2693(86)90293-5} {\bibfield
  {journal} {\bibinfo  {journal} {Phys. Lett. B}\ }\textbf {\bibinfo {volume}
  {180}},\ \bibinfo {pages} {185 } (\bibinfo {year} {1986})}\BibitemShut
  {NoStop}%
\bibitem [{\citenamefont {Freer}\ \emph {et~al.}(1995)\citenamefont {Freer},
  \citenamefont {Betts},\ and\ \citenamefont {Wuosmaa}}]{Freer95}%
  \BibitemOpen
  \bibfield  {author} {\bibinfo {author} {\bibfnamefont {M.}~\bibnamefont
  {Freer}}, \bibinfo {author} {\bibfnamefont {R.}~\bibnamefont {Betts}}, \ and\
  \bibinfo {author} {\bibfnamefont {A.}~\bibnamefont {Wuosmaa}},\ }\href
  {\doibase http://dx.doi.org/10.1016/0375-9474(94)00820-D} {\bibfield
  {journal} {\bibinfo  {journal} {Nucl. Phys. A}\ }\textbf {\bibinfo {volume}
  {587}},\ \bibinfo {pages} {36 } (\bibinfo {year} {1995})}\BibitemShut
  {NoStop}%
\bibitem [{\citenamefont {Maruhn}\ \emph {et~al.}(2010)\citenamefont {Maruhn},
  \citenamefont {Loebl}, \citenamefont {Itagaki},\ and\ \citenamefont
  {Kimura}}]{maruhn2010linear}%
  \BibitemOpen
  \bibfield  {author} {\bibinfo {author} {\bibfnamefont {J.}~\bibnamefont
  {Maruhn}}, \bibinfo {author} {\bibfnamefont {N.}~\bibnamefont {Loebl}},
  \bibinfo {author} {\bibfnamefont {N.}~\bibnamefont {Itagaki}}, \ and\
  \bibinfo {author} {\bibfnamefont {M.}~\bibnamefont {Kimura}},\ }\href@noop {}
  {\bibfield  {journal} {\bibinfo  {journal} {Nucl. Phys. A}\ }\textbf
  {\bibinfo {volume} {833}},\ \bibinfo {pages} {1} (\bibinfo {year}
  {2010})}\BibitemShut {NoStop}%
\bibitem [{\citenamefont {Ichikawa}\ \emph {et~al.}(2011)\citenamefont
  {Ichikawa}, \citenamefont {Maruhn}, \citenamefont {Itagaki},\ and\
  \citenamefont {Ohkubo}}]{Ichikawa11}%
  \BibitemOpen
  \bibfield  {author} {\bibinfo {author} {\bibfnamefont {T.}~\bibnamefont
  {Ichikawa}}, \bibinfo {author} {\bibfnamefont {J.~A.}\ \bibnamefont
  {Maruhn}}, \bibinfo {author} {\bibfnamefont {N.}~\bibnamefont {Itagaki}}, \
  and\ \bibinfo {author} {\bibfnamefont {S.}~\bibnamefont {Ohkubo}},\ }\href
  {\doibase 10.1103/PhysRevLett.107.112501} {\bibfield  {journal} {\bibinfo
  {journal} {Phys. Rev. Lett.}\ }\textbf {\bibinfo {volume} {107}},\ \bibinfo
  {pages} {112501} (\bibinfo {year} {2011})}\BibitemShut {NoStop}%
\bibitem [{\citenamefont {Ebran}\ \emph {et~al.}(2012)\citenamefont {Ebran},
  \citenamefont {Khan}, \citenamefont {Nik{\v{s}}i{\'c}},\ and\ \citenamefont
  {Vretenar}}]{ebran2012atomic}%
  \BibitemOpen
  \bibfield  {author} {\bibinfo {author} {\bibfnamefont {J.-P.}\ \bibnamefont
  {Ebran}}, \bibinfo {author} {\bibfnamefont {E.}~\bibnamefont {Khan}},
  \bibinfo {author} {\bibfnamefont {T.}~\bibnamefont {Nik{\v{s}}i{\'c}}}, \
  and\ \bibinfo {author} {\bibfnamefont {D.}~\bibnamefont {Vretenar}},\
  }\href@noop {} {\bibfield  {journal} {\bibinfo  {journal} {Nature}\ }\textbf
  {\bibinfo {volume} {487}},\ \bibinfo {pages} {341} (\bibinfo {year}
  {2012})}\BibitemShut {NoStop}%
\bibitem [{\citenamefont {Ebran}\ \emph
  {et~al.}(2014{\natexlab{b}})\citenamefont {Ebran}, \citenamefont {Khan},
  \citenamefont {Nik{\v{s}}i{\'c}},\ and\ \citenamefont
  {Vretenar}}]{ebran2014density}%
  \BibitemOpen
  \bibfield  {author} {\bibinfo {author} {\bibfnamefont {J.-P.}\ \bibnamefont
  {Ebran}}, \bibinfo {author} {\bibfnamefont {E.}~\bibnamefont {Khan}},
  \bibinfo {author} {\bibfnamefont {T.}~\bibnamefont {Nik{\v{s}}i{\'c}}}, \
  and\ \bibinfo {author} {\bibfnamefont {D.}~\bibnamefont {Vretenar}},\
  }\href@noop {} {\bibfield  {journal} {\bibinfo  {journal} {Phys. Rev. C}\
  }\textbf {\bibinfo {volume} {90}},\ \bibinfo {pages} {054329} (\bibinfo
  {year} {2014}{\natexlab{b}})}\BibitemShut {NoStop}%
\bibitem [{\citenamefont {Hecht}(1977)}]{Hecht77}%
  \BibitemOpen
  \bibfield  {author} {\bibinfo {author} {\bibfnamefont {K.~T.}\ \bibnamefont
  {Hecht}},\ }\href {\doibase 10.1103/PhysRevC.16.2401} {\bibfield  {journal}
  {\bibinfo  {journal} {Phys. Rev. C}\ }\textbf {\bibinfo {volume} {16}},\
  \bibinfo {pages} {2401} (\bibinfo {year} {1977})}\BibitemShut {NoStop}%
\bibitem [{\citenamefont {Nazarewicz}\ and\ \citenamefont
  {Dobaczewski}(1992)}]{Nazarewicz1992}%
  \BibitemOpen
  \bibfield  {author} {\bibinfo {author} {\bibfnamefont {W.}~\bibnamefont
  {Nazarewicz}}\ and\ \bibinfo {author} {\bibfnamefont {J.}~\bibnamefont
  {Dobaczewski}},\ }\href
  {http://journals.aps.org/prl/abstract/10.1103/PhysRevLett.68.154} {\bibfield
  {journal} {\bibinfo  {journal} {Phys. Rev. Lett.}\ }\textbf {\bibinfo
  {volume} {68}},\ \bibinfo {pages} {154} (\bibinfo {year} {1992})}\BibitemShut
  {NoStop}%
\bibitem [{\citenamefont {Becke}\ and\ \citenamefont
  {Edgecombe}(1990)}]{Becke1990}%
  \BibitemOpen
  \bibfield  {author} {\bibinfo {author} {\bibfnamefont {A.~D.}\ \bibnamefont
  {Becke}}\ and\ \bibinfo {author} {\bibfnamefont {K.~E.}\ \bibnamefont
  {Edgecombe}},\ }\href {\doibase http://dx.doi.org/10.1063/1.458517}
  {\bibfield  {journal} {\bibinfo  {journal} {J. Phys. Chem.}\ }\textbf
  {\bibinfo {volume} {92}},\ \bibinfo {pages} {5397} (\bibinfo {year}
  {1990})}\BibitemShut {NoStop}%
\bibitem [{\citenamefont {Savin}\ \emph {et~al.}(1997)\citenamefont {Savin},
  \citenamefont {Nesper}, \citenamefont {Wengert},\ and\ \citenamefont
  {F{\"a}ssler}}]{savin1997elf}%
  \BibitemOpen
  \bibfield  {author} {\bibinfo {author} {\bibfnamefont {A.}~\bibnamefont
  {Savin}}, \bibinfo {author} {\bibfnamefont {R.}~\bibnamefont {Nesper}},
  \bibinfo {author} {\bibfnamefont {S.}~\bibnamefont {Wengert}}, \ and\
  \bibinfo {author} {\bibfnamefont {T.~F.}\ \bibnamefont {F{\"a}ssler}},\
  }\href@noop {} {\bibfield  {journal} {\bibinfo  {journal} {Angew. Chem. Int.
  Ed. Engl.}\ }\textbf {\bibinfo {volume} {36}},\ \bibinfo {pages} {1808}
  (\bibinfo {year} {1997})}\BibitemShut {NoStop}%
\bibitem [{\citenamefont {Scemama}\ \emph {et~al.}(2004)\citenamefont
  {Scemama}, \citenamefont {Chaquin},\ and\ \citenamefont
  {Caffarel}}]{scemama2004electron}%
  \BibitemOpen
  \bibfield  {author} {\bibinfo {author} {\bibfnamefont {A.}~\bibnamefont
  {Scemama}}, \bibinfo {author} {\bibfnamefont {P.}~\bibnamefont {Chaquin}}, \
  and\ \bibinfo {author} {\bibfnamefont {M.}~\bibnamefont {Caffarel}},\
  }\href@noop {} {\bibfield  {journal} {\bibinfo  {journal} {J. Chem. Phys.}\
  }\textbf {\bibinfo {volume} {121}},\ \bibinfo {pages} {1725} (\bibinfo {year}
  {2004})}\BibitemShut {NoStop}%
\bibitem [{\citenamefont {Kohout}(2004)}]{Kohout04}%
  \BibitemOpen
  \bibfield  {author} {\bibinfo {author} {\bibfnamefont {M.}~\bibnamefont
  {Kohout}},\ }\href {\doibase 10.1002/qua.10768} {\bibfield  {journal}
  {\bibinfo  {journal} {Int. J. Quantum Chem.}\ }\textbf {\bibinfo {volume}
  {97}},\ \bibinfo {pages} {651} (\bibinfo {year} {2004})}\BibitemShut
  {NoStop}%
\bibitem [{\citenamefont {Burnus}\ \emph {et~al.}(2005)\citenamefont {Burnus},
  \citenamefont {Marques},\ and\ \citenamefont {Gross}}]{burnus2005time}%
  \BibitemOpen
  \bibfield  {author} {\bibinfo {author} {\bibfnamefont {T.}~\bibnamefont
  {Burnus}}, \bibinfo {author} {\bibfnamefont {M.~A.}\ \bibnamefont {Marques}},
  \ and\ \bibinfo {author} {\bibfnamefont {E.~K.}\ \bibnamefont {Gross}},\
  }\href@noop {} {\bibfield  {journal} {\bibinfo  {journal} {Phys. Rev. A}\
  }\textbf {\bibinfo {volume} {71}},\ \bibinfo {pages} {010501} (\bibinfo
  {year} {2005})}\BibitemShut {NoStop}%
\bibitem [{\citenamefont {Poater}\ \emph {et~al.}(2005)\citenamefont {Poater},
  \citenamefont {M.}, \citenamefont {{Sol{\`a}}},\ and\ \citenamefont
  {Silvi}}]{Poater}%
  \BibitemOpen
  \bibfield  {author} {\bibinfo {author} {\bibfnamefont {J.}~\bibnamefont
  {Poater}}, \bibinfo {author} {\bibfnamefont {D.}~\bibnamefont {M.}}, \bibinfo
  {author} {\bibfnamefont {M.}~\bibnamefont {{Sol{\`a}}}}, \ and\ \bibinfo
  {author} {\bibfnamefont {B.}~\bibnamefont {Silvi}},\ }\href {\doibase
  10.1021/cr030085x} {\bibfield  {journal} {\bibinfo  {journal} {Chem. Rev.}\
  }\textbf {\bibinfo {volume} {105}},\ \bibinfo {pages} {3911} (\bibinfo {year}
  {2005})}\BibitemShut {NoStop}%
\bibitem [{\citenamefont {Reinhard}\ \emph {et~al.}(2011)\citenamefont
  {Reinhard}, \citenamefont {Maruhn}, \citenamefont {Umar},\ and\ \citenamefont
  {Oberacker}}]{Reinhard2011}%
  \BibitemOpen
  \bibfield  {author} {\bibinfo {author} {\bibfnamefont {P.-G.}\ \bibnamefont
  {Reinhard}}, \bibinfo {author} {\bibfnamefont {J.~A.}\ \bibnamefont
  {Maruhn}}, \bibinfo {author} {\bibfnamefont {A.~S.}\ \bibnamefont {Umar}}, \
  and\ \bibinfo {author} {\bibfnamefont {V.~E.}\ \bibnamefont {Oberacker}},\
  }\href {\doibase 10.1103/PhysRevC.83.034312} {\bibfield  {journal} {\bibinfo
  {journal} {Phys. Rev. C}\ }\textbf {\bibinfo {volume} {83}},\ \bibinfo
  {pages} {034312} (\bibinfo {year} {2011})}\BibitemShut {NoStop}%
\bibitem [{\citenamefont {Ring}\ and\ \citenamefont
  {Schuck}(2004)}]{ring2004nuclear}%
  \BibitemOpen
  \bibfield  {author} {\bibinfo {author} {\bibfnamefont {P.}~\bibnamefont
  {Ring}}\ and\ \bibinfo {author} {\bibfnamefont {P.}~\bibnamefont {Schuck}},\
  }\href@noop {} {\emph {\bibinfo {title} {The nuclear many-body problem}}}\
  (\bibinfo  {publisher} {Springer Science \& Business Media},\ \bibinfo {year}
  {2004})\BibitemShut {NoStop}%
\bibitem [{\citenamefont {Bogner}\ \emph {et~al.}(2013)\citenamefont {Bogner},
  \citenamefont {Bulgac}, \citenamefont {Carlson}, \citenamefont {Engel},
  \citenamefont {Fann}, \citenamefont {Furnstahl}, \citenamefont {Gandolfi},
  \citenamefont {Hagen}, \citenamefont {Horoi},\ and\ \citenamefont
  {Johnson}}]{bogner2013}%
  \BibitemOpen
  \bibfield  {author} {\bibinfo {author} {\bibfnamefont {S.}~\bibnamefont
  {Bogner}}, \bibinfo {author} {\bibfnamefont {A.}~\bibnamefont {Bulgac}},
  \bibinfo {author} {\bibfnamefont {J.}~\bibnamefont {Carlson}}, \bibinfo
  {author} {\bibfnamefont {J.}~\bibnamefont {Engel}}, \bibinfo {author}
  {\bibfnamefont {G.}~\bibnamefont {Fann}}, \bibinfo {author} {\bibfnamefont
  {R.~J.}\ \bibnamefont {Furnstahl}}, \bibinfo {author} {\bibfnamefont
  {S.}~\bibnamefont {Gandolfi}}, \bibinfo {author} {\bibfnamefont
  {G.}~\bibnamefont {Hagen}}, \bibinfo {author} {\bibfnamefont
  {M.}~\bibnamefont {Horoi}}, \ and\ \bibinfo {author} {\bibfnamefont
  {C.}~\bibnamefont {Johnson}},\ }\href@noop {} {\bibfield  {journal} {\bibinfo
   {journal} {Comput. Phys. Comm.}\ }\textbf {\bibinfo {volume} {184}},\
  \bibinfo {pages} {2235} (\bibinfo {year} {2013})}\BibitemShut {NoStop}%
\bibitem [{\citenamefont {Erler}\ \emph {et~al.}(2012)\citenamefont {Erler},
  \citenamefont {Birge}, \citenamefont {Kortelainen}, \citenamefont
  {Nazarewicz}, \citenamefont {Olsen}, \citenamefont {Perhac},\ and\
  \citenamefont {Stoitsov}}]{erler12}%
  \BibitemOpen
  \bibfield  {author} {\bibinfo {author} {\bibfnamefont {J.}~\bibnamefont
  {Erler}}, \bibinfo {author} {\bibfnamefont {N.}~\bibnamefont {Birge}},
  \bibinfo {author} {\bibfnamefont {M.}~\bibnamefont {Kortelainen}}, \bibinfo
  {author} {\bibfnamefont {W.}~\bibnamefont {Nazarewicz}}, \bibinfo {author}
  {\bibfnamefont {E.}~\bibnamefont {Olsen}}, \bibinfo {author} {\bibfnamefont
  {A.~M.}\ \bibnamefont {Perhac}}, \ and\ \bibinfo {author} {\bibfnamefont
  {M.}~\bibnamefont {Stoitsov}},\ }\href
  {http://www.nature.com/nature/journal/v486/n7404/full/nature11188.html}
  {\bibfield  {journal} {\bibinfo  {journal} {Nature}\ }\textbf {\bibinfo
  {volume} {486}},\ \bibinfo {pages} {509} (\bibinfo {year}
  {2012})}\BibitemShut {NoStop}%
\bibitem [{\citenamefont {Agbemava}\ \emph {et~al.}(2014)\citenamefont
  {Agbemava}, \citenamefont {Afanasjev}, \citenamefont {Ray},\ and\
  \citenamefont {Ring}}]{Agbemava14}%
  \BibitemOpen
  \bibfield  {author} {\bibinfo {author} {\bibfnamefont {S.~E.}\ \bibnamefont
  {Agbemava}}, \bibinfo {author} {\bibfnamefont {A.~V.}\ \bibnamefont
  {Afanasjev}}, \bibinfo {author} {\bibfnamefont {D.}~\bibnamefont {Ray}}, \
  and\ \bibinfo {author} {\bibfnamefont {P.}~\bibnamefont {Ring}},\ }\href
  {\doibase 10.1103/PhysRevC.89.054320} {\bibfield  {journal} {\bibinfo
  {journal} {Phys. Rev. C}\ }\textbf {\bibinfo {volume} {89}},\ \bibinfo
  {pages} {054320} (\bibinfo {year} {2014})}\BibitemShut {NoStop}%
\bibitem [{\citenamefont {Kortelainen}\ \emph {et~al.}(2012)\citenamefont
  {Kortelainen}, \citenamefont {McDonnell}, \citenamefont {Nazarewicz},
  \citenamefont {Reinhard}, \citenamefont {Sarich}, \citenamefont {Schunck},
  \citenamefont {Stoitsov},\ and\ \citenamefont
  {Wild}}]{kortelainen2012nuclear}%
  \BibitemOpen
  \bibfield  {author} {\bibinfo {author} {\bibfnamefont {M.}~\bibnamefont
  {Kortelainen}}, \bibinfo {author} {\bibfnamefont {J.}~\bibnamefont
  {McDonnell}}, \bibinfo {author} {\bibfnamefont {W.}~\bibnamefont
  {Nazarewicz}}, \bibinfo {author} {\bibfnamefont {P.-G.}\ \bibnamefont
  {Reinhard}}, \bibinfo {author} {\bibfnamefont {J.}~\bibnamefont {Sarich}},
  \bibinfo {author} {\bibfnamefont {N.}~\bibnamefont {Schunck}}, \bibinfo
  {author} {\bibfnamefont {M.}~\bibnamefont {Stoitsov}}, \ and\ \bibinfo
  {author} {\bibfnamefont {S.}~\bibnamefont {Wild}},\ }\href@noop {} {\bibfield
   {journal} {\bibinfo  {journal} {Phys. Rev. C}\ }\textbf {\bibinfo {volume}
  {85}},\ \bibinfo {pages} {024304} (\bibinfo {year} {2012})}\BibitemShut
  {NoStop}%
\bibitem [{\citenamefont {Stoitsov}\ \emph {et~al.}(2003)\citenamefont
  {Stoitsov}, \citenamefont {Dobaczewski}, \citenamefont {Nazarewicz},
  \citenamefont {Pittel},\ and\ \citenamefont {Dean}}]{Sto03}%
  \BibitemOpen
  \bibfield  {author} {\bibinfo {author} {\bibfnamefont {M.~V.}\ \bibnamefont
  {Stoitsov}}, \bibinfo {author} {\bibfnamefont {J.}~\bibnamefont
  {Dobaczewski}}, \bibinfo {author} {\bibfnamefont {W.}~\bibnamefont
  {Nazarewicz}}, \bibinfo {author} {\bibfnamefont {S.}~\bibnamefont {Pittel}},
  \ and\ \bibinfo {author} {\bibfnamefont {D.~J.}\ \bibnamefont {Dean}},\
  }\href {\doibase 10.1103/PhysRevC.68.054312} {\bibfield  {journal} {\bibinfo
  {journal} {Phys. Rev. C}\ }\textbf {\bibinfo {volume} {68}},\ \bibinfo
  {pages} {054312} (\bibinfo {year} {2003})}\BibitemShut {NoStop}%
\bibitem [{\citenamefont {Schunck}\ \emph {et~al.}(2012)\citenamefont
  {Schunck}, \citenamefont {Dobaczewski}, \citenamefont {McDonnell},
  \citenamefont {Satu{\l}a}, \citenamefont {Sheikh}, \citenamefont {Staszczak},
  \citenamefont {Stoitsov},\ and\ \citenamefont
  {Toivanen}}]{schunck2012solution}%
  \BibitemOpen
  \bibfield  {author} {\bibinfo {author} {\bibfnamefont {N.}~\bibnamefont
  {Schunck}}, \bibinfo {author} {\bibfnamefont {J.}~\bibnamefont
  {Dobaczewski}}, \bibinfo {author} {\bibfnamefont {J.}~\bibnamefont
  {McDonnell}}, \bibinfo {author} {\bibfnamefont {W.}~\bibnamefont
  {Satu{\l}a}}, \bibinfo {author} {\bibfnamefont {J.}~\bibnamefont {Sheikh}},
  \bibinfo {author} {\bibfnamefont {A.}~\bibnamefont {Staszczak}}, \bibinfo
  {author} {\bibfnamefont {M.}~\bibnamefont {Stoitsov}}, \ and\ \bibinfo
  {author} {\bibfnamefont {P.}~\bibnamefont {Toivanen}},\ }\href@noop {}
  {\bibfield  {journal} {\bibinfo  {journal} {Comput. Phys. Commun.}\ }\textbf
  {\bibinfo {volume} {1}},\ \bibinfo {pages} {166} (\bibinfo {year}
  {2012})}\BibitemShut {NoStop}%
\bibitem [{\citenamefont {Engel}\ \emph {et~al.}(1975)\citenamefont {Engel},
  \citenamefont {Brink}, \citenamefont {Goeke}, \citenamefont {Krieger},\ and\
  \citenamefont {Vautherin}}]{Engel75}%
  \BibitemOpen
  \bibfield  {author} {\bibinfo {author} {\bibfnamefont {Y.}~\bibnamefont
  {Engel}}, \bibinfo {author} {\bibfnamefont {D.}~\bibnamefont {Brink}},
  \bibinfo {author} {\bibfnamefont {K.}~\bibnamefont {Goeke}}, \bibinfo
  {author} {\bibfnamefont {S.}~\bibnamefont {Krieger}}, \ and\ \bibinfo
  {author} {\bibfnamefont {D.}~\bibnamefont {Vautherin}},\ }\href {\doibase
  http://dx.doi.org/10.1016/0375-9474(75)90184-0} {\bibfield  {journal}
  {\bibinfo  {journal} {Nucl. Phys. A}\ }\textbf {\bibinfo {volume} {249}},\
  \bibinfo {pages} {215} (\bibinfo {year} {1975})}\BibitemShut {NoStop}%
\bibitem [{\citenamefont {Staszczak}\ \emph {et~al.}(2009)\citenamefont
  {Staszczak}, \citenamefont {Baran}, \citenamefont {Dobaczewski},\ and\
  \citenamefont {Nazarewicz}}]{Staszczak09}%
  \BibitemOpen
  \bibfield  {author} {\bibinfo {author} {\bibfnamefont {A.}~\bibnamefont
  {Staszczak}}, \bibinfo {author} {\bibfnamefont {A.}~\bibnamefont {Baran}},
  \bibinfo {author} {\bibfnamefont {J.}~\bibnamefont {Dobaczewski}}, \ and\
  \bibinfo {author} {\bibfnamefont {W.}~\bibnamefont {Nazarewicz}},\ }\href
  {\doibase 10.1103/PhysRevC.80.014309} {\bibfield  {journal} {\bibinfo
  {journal} {Phys. Rev. C}\ }\textbf {\bibinfo {volume} {80}},\ \bibinfo
  {pages} {014309} (\bibinfo {year} {2009})}\BibitemShut {NoStop}%
\bibitem [{\citenamefont {Sadhukhan}\ \emph {et~al.}(2014)\citenamefont
  {Sadhukhan}, \citenamefont {Dobaczewski}, \citenamefont {Nazarewicz},
  \citenamefont {Sheikh},\ and\ \citenamefont {Baran}}]{Sadhukhan14}%
  \BibitemOpen
  \bibfield  {author} {\bibinfo {author} {\bibfnamefont {J.}~\bibnamefont
  {Sadhukhan}}, \bibinfo {author} {\bibfnamefont {J.}~\bibnamefont
  {Dobaczewski}}, \bibinfo {author} {\bibfnamefont {W.}~\bibnamefont
  {Nazarewicz}}, \bibinfo {author} {\bibfnamefont {J.~A.}\ \bibnamefont
  {Sheikh}}, \ and\ \bibinfo {author} {\bibfnamefont {A.}~\bibnamefont
  {Baran}},\ }\href {\doibase 10.1103/PhysRevC.90.061304} {\bibfield  {journal}
  {\bibinfo  {journal} {Phys. Rev. C}\ }\textbf {\bibinfo {volume} {90}},\
  \bibinfo {pages} {061304} (\bibinfo {year} {2014})}\BibitemShut {NoStop}%
\bibitem [{\citenamefont {Simenel}\ and\ \citenamefont
  {Umar}(2014)}]{Simenel14}%
  \BibitemOpen
  \bibfield  {author} {\bibinfo {author} {\bibfnamefont {C.}~\bibnamefont
  {Simenel}}\ and\ \bibinfo {author} {\bibfnamefont {A.~S.}\ \bibnamefont
  {Umar}},\ }\href {\doibase 10.1103/PhysRevC.89.031601} {\bibfield  {journal}
  {\bibinfo  {journal} {Phys. Rev. C}\ }\textbf {\bibinfo {volume} {89}},\
  \bibinfo {pages} {031601} (\bibinfo {year} {2014})}\BibitemShut {NoStop}%
\bibitem [{\citenamefont {Zhao}\ \emph {et~al.}(2015)\citenamefont {Zhao},
  \citenamefont {Lu}, \citenamefont {Nik{\v{s}}i{\'{c}}},\ and\ \citenamefont
  {Vretenar}}]{Zhao15}%
  \BibitemOpen
  \bibfield  {author} {\bibinfo {author} {\bibfnamefont {J.}~\bibnamefont
  {Zhao}}, \bibinfo {author} {\bibfnamefont {B.-N.}\ \bibnamefont {Lu}},
  \bibinfo {author} {\bibfnamefont {T.}~\bibnamefont {Nik{\v{s}}i{\'{c}}}}, \
  and\ \bibinfo {author} {\bibfnamefont {D.}~\bibnamefont {Vretenar}},\ }\href
  {\doibase 10.1103/PhysRevC.92.064315} {\bibfield  {journal} {\bibinfo
  {journal} {Phys. Rev. C}\ }\textbf {\bibinfo {volume} {92}},\ \bibinfo
  {pages} {064315} (\bibinfo {year} {2015})}\BibitemShut {NoStop}%
\bibitem [{\citenamefont {Hulet}\ \emph {et~al.}(1989)\citenamefont {Hulet},
  \citenamefont {Wild}, \citenamefont {Dougan}, \citenamefont {Lougheed},
  \citenamefont {Landrum}, \citenamefont {Dougan}, \citenamefont {Baisden},
  \citenamefont {Henderson}, \citenamefont {Dupzyk}, \citenamefont {Hahn},
  \citenamefont {Sch\"adel}, \citenamefont {S\"ummerer},\ and\ \citenamefont
  {Bethune}}]{Hulet89}%
  \BibitemOpen
  \bibfield  {author} {\bibinfo {author} {\bibfnamefont {E.~K.}\ \bibnamefont
  {Hulet}}, \bibinfo {author} {\bibfnamefont {J.~F.}\ \bibnamefont {Wild}},
  \bibinfo {author} {\bibfnamefont {R.~J.}\ \bibnamefont {Dougan}}, \bibinfo
  {author} {\bibfnamefont {R.~W.}\ \bibnamefont {Lougheed}}, \bibinfo {author}
  {\bibfnamefont {J.~H.}\ \bibnamefont {Landrum}}, \bibinfo {author}
  {\bibfnamefont {A.~D.}\ \bibnamefont {Dougan}}, \bibinfo {author}
  {\bibfnamefont {P.~A.}\ \bibnamefont {Baisden}}, \bibinfo {author}
  {\bibfnamefont {C.~M.}\ \bibnamefont {Henderson}}, \bibinfo {author}
  {\bibfnamefont {R.~J.}\ \bibnamefont {Dupzyk}}, \bibinfo {author}
  {\bibfnamefont {R.~L.}\ \bibnamefont {Hahn}}, \bibinfo {author}
  {\bibfnamefont {M.}~\bibnamefont {Sch\"adel}}, \bibinfo {author}
  {\bibfnamefont {K.}~\bibnamefont {S\"ummerer}}, \ and\ \bibinfo {author}
  {\bibfnamefont {G.~R.}\ \bibnamefont {Bethune}},\ }\href {\doibase
  10.1103/PhysRevC.40.770} {\bibfield  {journal} {\bibinfo  {journal} {Phys.
  Rev. C}\ }\textbf {\bibinfo {volume} {40}},\ \bibinfo {pages} {770} (\bibinfo
  {year} {1989})}\BibitemShut {NoStop}%
\bibitem [{\citenamefont {McDonnell}\ \emph {et~al.}(2013)\citenamefont
  {McDonnell}, \citenamefont {Nazarewicz},\ and\ \citenamefont
  {Sheikh}}]{McDonnell2013}%
  \BibitemOpen
  \bibfield  {author} {\bibinfo {author} {\bibfnamefont {J.~D.}\ \bibnamefont
  {McDonnell}}, \bibinfo {author} {\bibfnamefont {W.}~\bibnamefont
  {Nazarewicz}}, \ and\ \bibinfo {author} {\bibfnamefont {J.~A.}\ \bibnamefont
  {Sheikh}},\ }\href {\doibase 10.1103/PhysRevC.87.054327} {\bibfield
  {journal} {\bibinfo  {journal} {Phys. Rev. C}\ }\textbf {\bibinfo {volume}
  {87}},\ \bibinfo {pages} {054327} (\bibinfo {year} {2013})}\BibitemShut
  {NoStop}%
\bibitem [{\citenamefont {Sadhukhan}\ \emph {et~al.}(2016)\citenamefont
  {Sadhukhan}, \citenamefont {Nazarewicz},\ and\ \citenamefont
  {Schunck}}]{Sadhukhan2016}%
  \BibitemOpen
  \bibfield  {author} {\bibinfo {author} {\bibfnamefont {J.}~\bibnamefont
  {Sadhukhan}}, \bibinfo {author} {\bibfnamefont {W.}~\bibnamefont
  {Nazarewicz}}, \ and\ \bibinfo {author} {\bibfnamefont {N.}~\bibnamefont
  {Schunck}},\ }\href {\doibase 10.1103/PhysRevC.93.011304} {\bibfield
  {journal} {\bibinfo  {journal} {Phys. Rev. C}\ }\textbf {\bibinfo {volume}
  {93}},\ \bibinfo {pages} {011304} (\bibinfo {year} {2016})}\BibitemShut
  {NoStop}%
\end{thebibliography}

%

%
\end{document}